\documentclass[usenatbib]{mn2e} \usepackage{psfig}

\title[Roche tomography of cataclysmic variables -- III. Starspots on
AE Aqr] {Roche tomography of cataclysmic variables -- III. Starspots
on AE Aqr}

\author[C.\,A.\ Watson, V.\,S.\ Dhillon and T.\ Shahbaz]
{C.\,A.\ Watson,$^1$\thanks{E-mail: c.watson@sheffield.ac.uk}
V.\,S.\ Dhillon,$^1$ and T.\ Shahbaz$^2$ \\
$^1$ Department of Physics and Astronomy, University of Sheffield,
Sheffield S3 7RH, UK\\ $^2$ Instituto de Astrof\'{i}sica de Canarias,
38200 La Laguna, Tenerife, Spain\\}

\date{\center{\Large Submitted for publication in the Monthly
Notices of the Royal Astronomical Society \\
\vspace{.5cm} \today}}

\defcitealias{barnes01a}{Barnes et al. 2001a}
\defcitealias{barnes01b}{Barnes et al. 2001b}

\begin{document}
\maketitle

\begin{abstract}

We present a Roche tomography reconstruction of the secondary star in
the cataclysmic variable AE Aqr. The tomogram reveals several
surface inhomogeneities that are due to the presence of large, cool
starspots. In addition to a number of lower-latitude spots, the maps
also show the presence of a large high latitude spot similar to that
seen in Doppler images of rapidly-rotating isolated stars, and a
relative paucity of spots at a latitude of 40$^{\circ}$. In total, we
estimate that some 18 per cent of the Northern hemisphere of AE Aqr is
spotted.

We have also applied the {\em entropy landscape} technique to
determine accurate parameters for the binary system. We obtain optimal
masses of $M_1$ = 0.74 M$_{\odot}$, $M_2$ = 0.50 M$_{\odot}$, a
systemic velocity $\gamma$ = --63 km s$^{-1}$ and an orbital
inclination of $i = 66^{\circ}$.

Given that this is the first study to successfully image starspots on
the secondary star in a cataclysmic variable, we discuss the role that
further studies of this kind may play in our understanding of these
binaries.

\end{abstract}


\section{Introduction}
\label{sec:intro}

Cataclysmic variables (CVs) are short period binary systems in which a
(typically) late main-sequence star (the secondary) transfers material
via Roche-lobe overflow to a white dwarf primary star. Although CVs
are largely observed to study the fundamental astrophysical process of
accretion, it is the Roche-lobe filling secondary stars themselves
that are key to our understanding of the origin, evolution and
behaviour of this class of interacting binary.

For instance, magnetic activity of the secondary star plays a pivotal
role in the canonical theory of CV evolution by draining angular
momentum from the binary through a process known as {\em magnetic
braking} (e.g. \citealt{kraft67}, \citealt{mestel68},
\citealt{spruit83}).  This is believed to be responsible for
sustaining mass transfer between the components and to cause CVs to
evolve to shorter orbital periods. At a secondary star mass of $\sim$
0.25 $M_{\odot}$, corresponding to an orbital period of around 3
hours, the secondary star becomes fully convective. This is thought to
quench the dynamo mechanism and shut down the mass transfer.  Contact
is later re-established at a shorter orbital period ($\sim$ 2 hours)
due to angular momentum loss by gravitational radiation -- explaining
the dearth of CVs seen between 2--3 hours (the so-called period gap --
see e.g. \citealt{taam89}). Although this is the canonical picture of
CV evolution, \citet{andronov03} have noted that it is unclear if
magnetic braking is the dominant angular momentum loss above the
period gap, and there is no observational support for an abrupt change
in angular momentum losses at the point where stars become fully
convective.

Furthermore, magnetic activity cycles on the secondary star have been
invoked to explain the orbital period variations observed for some CVs
\citep{applegate92}. In addition, an increase in the number of
magnetic flux tubes on the secondary star may cause the star to expand
relative to its Roche lobe (e.g. \citealt{richman94}) and result in
enhanced mass transfer -- giving rise to an increased mass transfer
rate through the accretion disc and a corresponding increase in the
system luminosity. The additional mass flow would also reduce the time
required to build up enough material in the disc to trigger an
outburst -- resulting in shorter time intervals between consecutive
outbursts.

It is also speculated that the secondary star magnetic field
could reach across to the disc and remove angular momentum from its
outer regions, causing material to accumulate in the inner disc.
An increase in the number of magnetic flux tubes on the secondary
could then cause material to accumulate in the inner regions faster 
(e.g. \citealt{ak01}) -- resulting in shorter outburst durations.
Thus, magnetic phenomena on the secondary star have been invoked to
explain variations in the orbital periods, mean brightnesses, mean
outburst frequencies, mean outburst durations and outburst shapes of
CVs, as well as playing a major role in their evolution.

From another (perhaps more important) perspective, by studying
magnetic activity in CVs we can provide tests of stellar dynamo
theories. Although faint ($m_v >$ 11), CVs contain some of the most
rapidly rotating stars known, and therefore provide an excellent
test-bed in which to study such models since they should have strong
dynamo action (e.g. \citealt{rutten87}). Furthermore, tidal forces are
thought to have a strong impact on stellar dynamos, suppressing
differential rotation \citep{scharlemann82} and forcing starspots to
form at preferred longitudes \citep{holzwarth03}. Currently,
observational studies of the impact of tidal effects on stellar
dynamos are still lacking.  Since the secondary stars in CVs are
heavily distorted, studies of magnetic activity on these stars are
highly desirable.

Despite the obvious importance of magnetic activity to our
understanding of the evolution and behaviour of these binaries, and
the role that studying such activity on the secondary stars in CVs can
play in furthering our understanding of the solar-stellar connection,
little work has been carried out in this field (with the notable
exception of the work by \citealt{webb02}). We aim to
address this issue. In this paper we present
the first images of starspots on the secondary star in a CV (AE Aqr)
using Roche tomography (see \citealt{watson01a}; \citealt{dhillon01}
and \citealt{watson03} for details).

\section{Observations and reduction}

Simultaneous spectroscopic and photometric data of AE Aqr were
acquired over two nights on 2001 August 9 and 10. The spectroscopic
data were obtained using the 4.2-m William Herschel Telescope (WHT)
and the simultaneous photometry was carried out using the 1.0-m
Jacobus Kapteyn Telescope (JKT). Both telescopes are part of the Isaac
Newton Group of telescopes situated on the island of La Palma.

\subsection{Spectroscopy}

The spectroscopic observations of AE Aqr were carried out using the
Utrecht Echelle Spectrograph (UES) and E31 grating on the WHT. The
SITe1 CCD chip with 2088 $\times$ 2120 pixels (each of 24$\mu$m) was
used. Centred at a wavelength of 5000\AA~in order 114, this allowed
orders 77--141 to be covered (corresponding to a wavelength range of
4001\AA~--7475\AA, with significant wavelength overlap between
adjacent orders). With a slit width of 1.11 arcsec (projecting onto
2.0 pixels at the detector) a spectral resolution of around 46000
(i.e. $\sim$6.6 km s$^{-1}$) was obtained. The majority of spectra
were taken using a 200-s exposure time in order to minimise velocity
smearing of the data due to the orbital motion of the secondary star.
Comparison ThAr arc lamp exposures were taken approximately every hour
to calibrate instrumental flexure.

With this setup we obtained 88 usable spectra (with one 400-s
exposure) on 2001 Aug 9, and 95 usable spectra on the following
night. Both nights were clear (though dusty) with seeing varying
between 1 and 1.4 arcsec at low air mass, worsening to 2 arcsec
towards the end of the second night. The signal-to-noise of each
spectrum peaked at $\sim$20--44 per pixel around
5580\AA. Table~\ref{table:log} gives a journal of the observations
undertaken at the WHT.

\begin{table*}
\caption[]{Log of the WHT spectroscopic observations of AE Aqr, the
spectral-type template stars, a telluric-correction star and the
spectrophotometric standard HR 26. The first column gives the object
name, columns 2--4 list the UT Date and the exposure start and end
times, respectively, columns 5--6 list the exposure times and number of
spectra taken for each object and the final column indicates the type
of science frame taken.}
\begin{tabular}{llccccl} \hline
Object & UT Date & UT Start & UT End & $T_{exp}$ (s) & No. spectra &
Comments \\ \hline AE Aqr & 2001 Aug 09 & 21:01 & 21:55 & 200 & 11 &
Target spectra \\ AE Aqr &  & 21:57 & 22:04 & 400 & 1 &  \\ AE Aqr & &
22:05 & 04:22 & 200 & 76 &  \\ HR 26 & 2001 Aug 10 & 04:29 & 04:29 &
10 & 1 & Telluric B9V star\\ HR 26 &  & 04:32 & 04:35 & 200 & 1 & \\
HR 26 &  & 04:37 & 04:42 & 300 & 1 &  \\ HD 19455 &  & 04:56 & 05:09 &
300 & 2 & Spectrophotometric standard \\ HR 9038 &  &  05:13 & 05:25 &
300 & 2 & K3V template \\ HR 8372 &  & 05:37 & 05:49 & 300 & 2 & K5V
template \\ HR 8881 &  & 05:53 & 06:05 & 300 & 2 & K1V template \\ AE
Aqr &  & 20:49 & 04:37 & 200 & 95 & Target spectra \\ HR 8086 & 2001
Aug 11 & 04:44 & 04:56 & 300 & 2 & K7V template \\ HR 8382 & & 05:04 &
05:16 & 300 & 2 & K2V template \\ HR 8 & & 05:20 & 05:31 & 300 & 2 &
K0V template \\ Gl 157A & & 05:42 & 06:14 & 300 & 4 & K4V Template \\
\hline
\label{table:log}
\end{tabular}
\end{table*}

\subsubsection{Data reduction}
\label{sec:spec_dr}

The raw data were first bias subtracted and the overscan regions
cropped. Pixel-to-pixel variations were corrected for using a tungsten
lamp flat-field. An order-by-order optimal extraction
(\citealt{horne86a}) was then carried out using the {\sc pamela}
package written by tom Marsh. The arc spectra were extracted from the
same locations on the detector as the target and fitted with a
fourth-order polynomial, which gave a typical rms scatter of better
than 0.001\AA. After extraction and wavelength calibration, each order
was then inspected visually for any obvious wavelength calibration
errors and to ensure that no jumps were present between adjacent
orders. Orders that were affected by bad columns on the SITe1 CCD chip
were then discarded (these were orders 113 and 120 centred on
5055\AA~and 4757\AA~respectively).

We should note here that Roche tomography cannot be performed on data
which has not been corrected for slit losses. This is because the
variable contribution of the secondary star to the total light of a CV
forces one to use relative line fluxes during the mapping process and
prohibits the usual method of normalising the spectra. We corrected
for slit losses by dividing each AE Aqr spectrum by the ratio of the
flux in the spectrum (after folding the spectrum through the
photometric filter response function) to the corresponding photometric
flux (see Section.~\ref{sec:photometry}).

\subsection{Photometry}
\label{sec:photometry}

Simultaneous photometric observations were carried out on the JKT
using a Kitt Peak V-band filter and the SITe2 CCD chip with
2088$\times$2120 pixels, each of 24$\mu$m. In order to reduce readout
time, 2 windows were created.  The first window (measuring
920$\times$1000 unbinned pixels) was setup to cover the target and the
brightest photometric standards from \citet{henden95}, and a second
window (measuring 920$\times$80 unbinned pixels) covered the bias
strip. The chip was then binned by a factor of 2$\times$2.

\subsubsection{Data reduction}

\begin{figure}
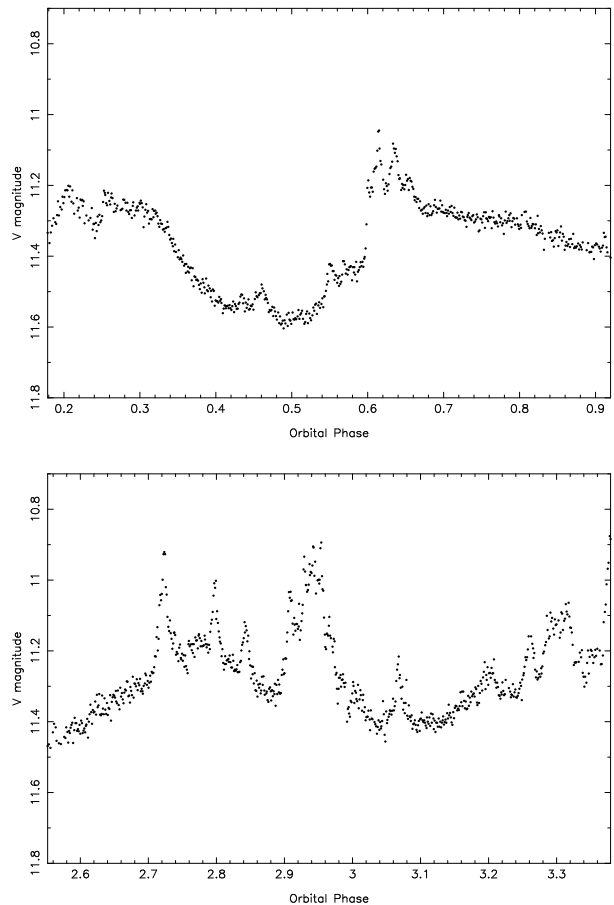

\psfig{figure=light1.ps,width=8.0cm,angle=-90.}
\psfig{figure=light2.ps,width=8.0cm,angle=-90.}
\caption{AE Aqr $V$-band JKT light curves. Top: data from 2001 August
  09.  Bottom: data from 2001 August 10. The data have been phased according
to the ephemeris derived in Section~\protect\ref{sec:ephemeris}.}
\label{fig:light}
\end{figure}

Since the master bias-frame showed that large scale structure in the
bias was insignificant, the bias level of the target frames was
removed by subtracting the median of the pixels in the overscan
region. Pixel-to-pixel variations were then removed by dividing the
target frames through by a master twilight flat-field frame. Aperture
photometry was performed using the package {\sc photom}. Finally, the
photometry data were corrected for atmospheric extinction by
subtracting the magnitude of a nearby comparison star (AE Aqr-2;
\citealt{henden95}). The light curves for both nights are shown in
Fig.~\ref{fig:light}.

Inspection of the light-curves clearly shows  the flaring activity so
typical of AE Aqr (e.g. \citealt{beskrovnaya96}; \citealt{skidmore03};
\citealt*{pearson03}), with flare amplitudes approaching $\sim$0.5
mag.

\subsection{Continuum fitting}

As mentioned earlier in Section~\ref{sec:spec_dr}, there is an unknown
contribution to each spectrum from the accretion regions. The
constantly changing continuum slope due to, for example, flaring (see
Fig~\ref{fig:light}) or from the varying aspect of the accretion
regions means that the construction of, for instance, a master
continuum fit to the data (such as that done by \citealt{cameron94})
is not appropriate under these circumstances. Furthermore, since the
contribution of the secondary star to the total light of the system is
constantly varying, normalisation of the continuum would also result
in the photospheric absorption lines from the secondary star varying
in relative strength from one exposure to the next. Therefore, we are
forced to subtract the continuum after each spectrum has been
corrected for slit-losses using simultaneous photometry
(Section~\ref{sec:photometry}).

The continuum fitting was carried out by placing spline knots at
locations that were relatively line-free. The positions of these
spline knots were chosen on an individual spectrum-by-spectrum basis
such that a smooth and visually acceptable fit was acquired, ignoring
regions that included emission lines. Naturally, this is a very
subjective procedure as it is particularly difficult to judge where
the true continuum actually lies amongst the forest of heavily
broadened lines.

In order to assess the impact of continuum mis-fits on our analysis we
experimented with different fits, including those that were quite
obviously incorrect. We were comforted to find that a poor continuum
fit did not adversely affect the shape of the line profiles after
carrying out the least squares deconvolution (LSD) process described
in Section~\ref{sec:lsd}.  Evidently the results of this process are
dominated by the sheer number of lines used rather than any continuum
mis-fits, even if these mis-fits amount to quite major departures from
the true continuum shape.

We did find, however, that we systematically fit the continuum at too
high a level, leading to LSD profiles with continuum (subtracted)
regions lying significantly below zero. This was rather unexpected as
the heavily broadened lines of the secondary act to `depress' the
observed continuum level, which would most likely lead to the
continuum being fitted at too low a level. We can only surmise that we
were successful in finding regions devoid of lines and that we fitted
across the top of the noise in these regions. This was easily solved,
however, by shifting the continuum fit to lower levels until the
continuum in the LSD profiles lay at zero. This did not affect the
shape of the line profiles obtained.

\section{Least squares deconvolution}
\label{sec:lsd}

Previous Roche tomograms have been restricted to using a single
absorption or emission line on a 4-m class telescope coupled with an
intermediate resolution spectrograph
(e.g. \citealt{watson03}). Although successful at picking out the
strong signatures of irradiation on the secondary stars, the features
due to starspots are far more subtle. In order to achieve high enough
signal-to-noise to detect starspot features in absorption-line
profiles of the secondary in AE Aqr we have applied the technique of
LSD.

LSD effectively stacks the $\sim$1000's of stellar absorption lines
observable in a single echelle spectrum to produce a single `average'
profile of greatly increased signal-to-noise. Theoretically, the
multiplex gain in signal-to-noise is the square root of the number of
lines observed. The technique was first applied to spectropolarimetric
observations of active stars by \cite{donati97a} and has since been
used in a large number of Doppler imaging studies by a variety of
groups (e.g. \citealt{barnes98}; \citealt{lister99}; \citealt{barnes00};
\citealt{barnes01}; \citetalias{barnes01a}; \citetalias{barnes01b};
\citealt{jeffers02}; \citealt{barnes04}; \citealt{marsden05}).
For supplementary details of the LSD technique, see also
\cite{cameron01} and \cite{barnes99}.

LSD assumes that starspots affect all of the rotationally broadened
profiles in the same manner. Although the amplitude of the starspot
bumps may vary from line to line, their morphology should be
identical.  LSD also requires the positions and relative strengths of
the lines observed in each echelle spectrum to be known. Currently, we
use line lists generated by the Vienna Atomic Line Database (VALD)
for this purpose (see {\citealt{kupka99}; \citealt{kupka00}).

The spectral type of AE Aqr has been determined to lie in the range
K3--K5 V/IV (\citealt{crawford56}; \citealt*{tanzi81}; \citealt{wade82};
\citealt{bruch91};  \citealt*{welsh93}; \citealt{reinsch94}; \citealt{
casares96}). Therefore, a line-list for a stellar atmosphere with
$T_{eff} = 4750K$ and $\log g = 4.5$ (the closest approximation
available in the database to a K4V spectral type) was downloaded for
use in the LSD process.  Since the line-list obtained from {\sc vald}
contain normalised line-depths, whereas Roche tomography uses
continuum {\em subtracted} spectra, each line-depth was scaled by a
fit to the continuum of a K4V template star. This meant that each
line's relative depth was now correct for use with continuum subtracted
data.

Although AE Aqr is slightly evolved (\citealt{eracleous96};
\citealt*{wynn97}; \citealt{pearson03}) and has non-solar abundances
due to possible CNO cycling (e.g. \citealt*{mauche97};
\citealt{schenker02}), our choice of line-list is unlikely to affect
the results presented in this paper. \cite{barnes99} investigated the
robustness of the LSD process with respect to the choice of the
line-list spectral type used.  By carrying out LSD on the rapidly
rotating single star PZ Tel (spectral type K0V) using different line
lists (from spectral type G2--K5), \cite{barnes99} found that the LSD
process was insensitive to the use of a line-list of incorrect
spectral type. He also concluded that, therefore, the effects of
metallicity would also have little impact on the LSD process.

One further complication in the LSD process that is possibly worth
considering is whether changes in the continuum lightcurve can result
in significant changes in the tilt of the spectrum. This would lead to
lines receiving a different weighting at different orbital phases which
may cause slight changes in the deconvolved profiles line-depths and
shift the effective `central wavelength' of the observations (which would
alter the effective limb darkening coefficient seen). These effects are
likely to be small, and certainly cannot be seen in this work.

We carried out LSD over the wavelength range of 4200\AA~-- 6820\AA.
Bluer than $\sim$4200\AA, the signal-to-noise is poor and degrades the
quality of the final deconvolved profile. For regions redder than
$\sim$6820\AA, the number of photospheric absorption lines is greatly
diminished and telluric features become more abundant -- inclusion of
which would degrade the final deconvolved profile. We also masked out
regions where strong emission lines were evident, such as around
H$\gamma$, H$\beta$ and H$\alpha$. Altogether, a total of 3843 lines
were used in the deconvolution process, resulting in an average
multiplex gain in signal-to-noise of 26 over single-line
studies. Errors were propagated through the LSD process, which results
in an under-estimate of the signal-to-noise ratio in the deconvolved
profiles (as also found by \cite{marsden05} -- see \citealt{wade00}
for more details). This results in our images being reconstructed to
reduced $\chi^2$ values less than 1, but does not affect the quality
of the final maps.

The individual deconvolved profiles obtained on 2001 August 9 and 10
are displayed in Figs.~\ref{fig:profs1} and ~\ref{fig:profs2},
respectively.  These clearly show bump features moving from blue (--ve
velocities) to red (+ve velocities) across the profile. These
features, as well as the orbital motion and variation in the projected
equatorial rotation velocity $v \sin i$, are also obvious when the LSD
profiles are trailed (see Fig~\ref{fig:trails}). In order to enhance
the spot features, we have subtracted a theoretical profile from each
LSD profile and removed the orbital motion -- the result is shown in
Fig.~\ref{fig:theor}. (The theoretical profile removed from the data
was generated using our Roche tomography code adopting the binary
parameters found for AE Aqr (see Section~\ref{sec:pars}), assuming a
featureless stellar surface and the same limb-darkening coefficients
as used in the final reconstructions).

\begin{figure*}
\psfig{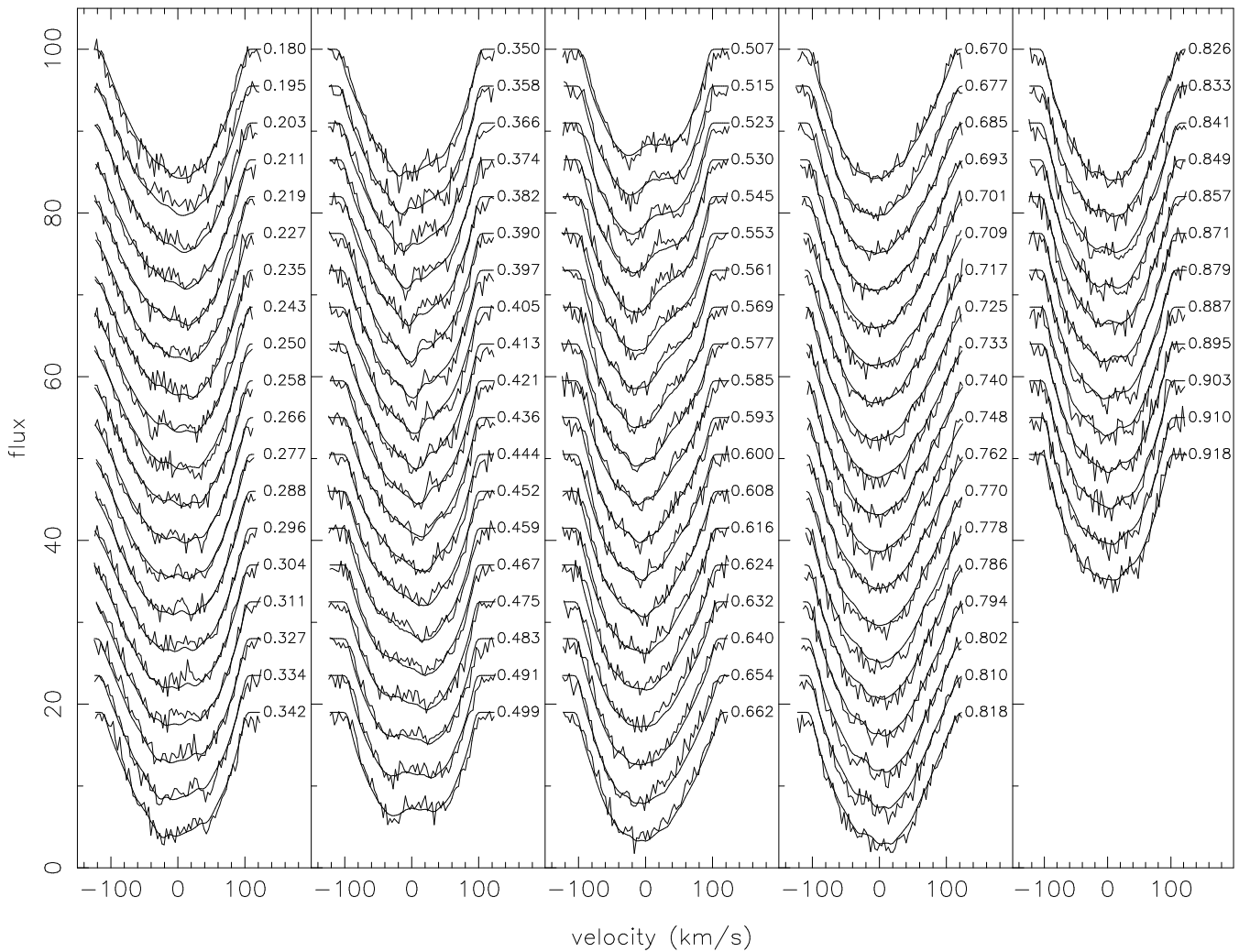}
\caption{The LSD profiles and maximum entropy fits (see
Section\protect~\ref{sec:maps} for further details) for AE Aqr for
observations starting on the night of 2001 August 9. The velocity of the
secondary star has been removed using the parameters found in
Section\protect~\ref{sec:pars}, and each profile is shifted vertically
for clarity. The orbital phase for each exposure is indicated at the
top right of each profile.}
\label{fig:profs1}
\end{figure*}

\begin{figure*}
\centering\psfig{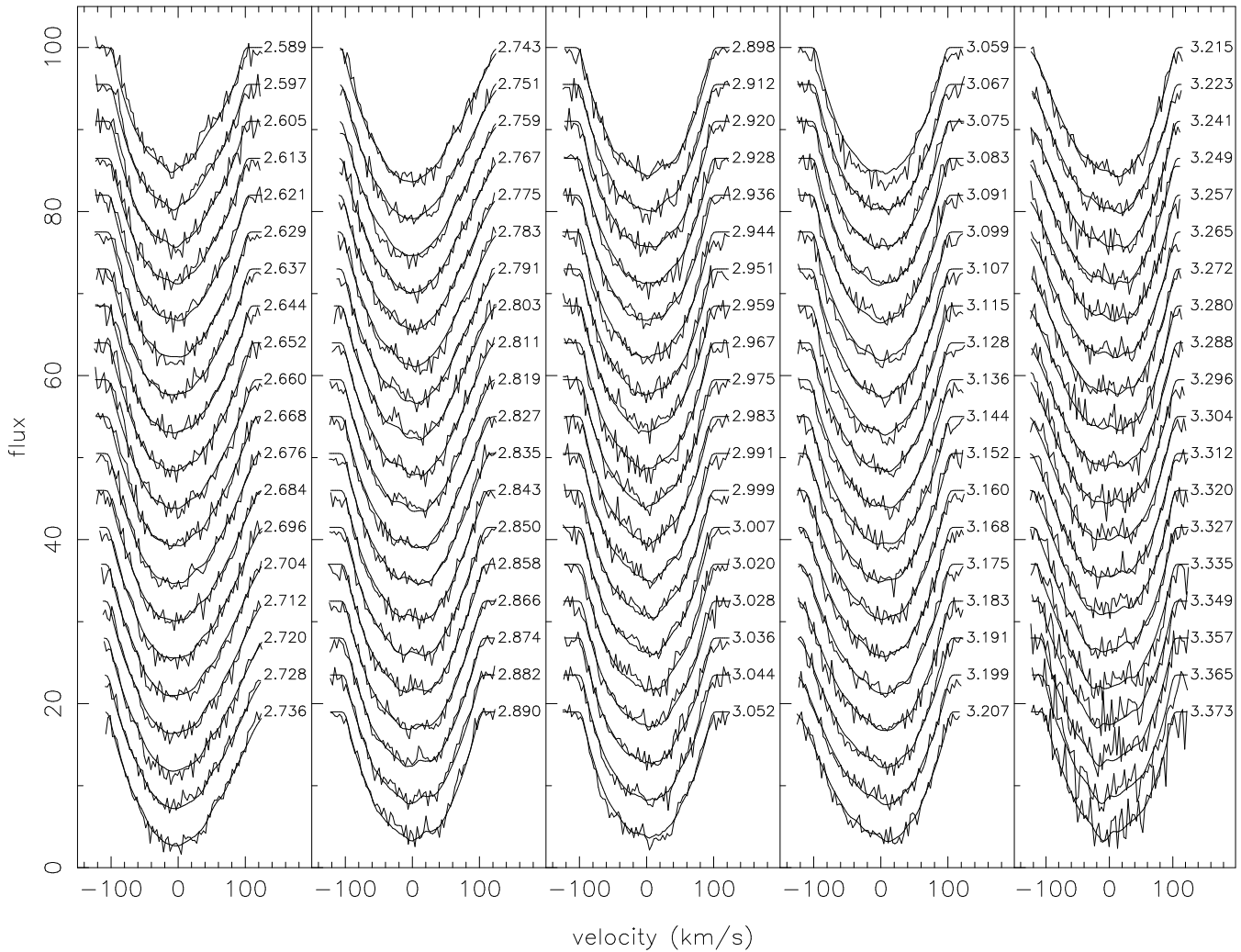}
\caption{The same as Fig.~\protect~\ref{fig:profs1} but for the second
night of WHT observations. The seeing deteriorated to $\sim$2 arcsec
towards the end of the night, hence the poorer signal-to-noise in the
latter profiles.}
\label{fig:profs2}
\end{figure*}

\begin{figure}
\psfig{figure=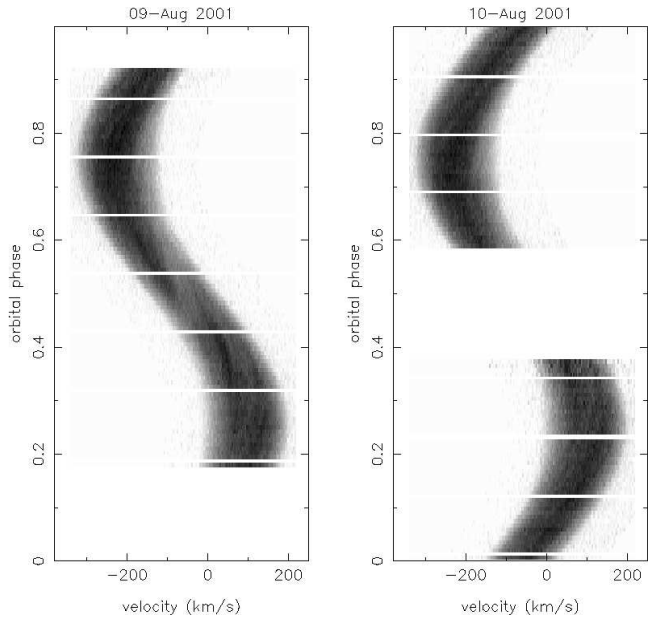,width=8.5cm,angle=-90.}
\caption{A trail of the deconvolved profiles from observations on 2001
August 9 (left) and 2001 August 10 (right). The small gaps in the phase
coverage are at times when arc spectra were taken for the purpose of
wavelength calibration. For clarity, gaps between individual exposures
due to readout time are not shown. In these trailed spectra, features
due to starspots will appear bright and several such features can be
seen which appear on both nights (see Fig.~\protect\ref{fig:theor} for
plots with increased contrast). Also evident is the variation in $v
\sin i$ showing the maximum at quadrature (phase 0.25 and 0.75) due to
the changing aspect of the tidally distorted star.}
\label{fig:trails}
\end{figure}

\begin{figure}
\psfig{figure=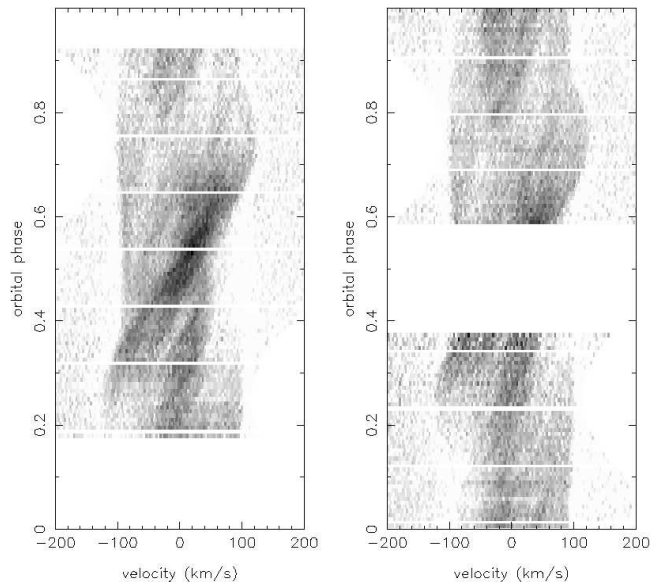,width=8.5cm,angle=-90.}
\caption{A trail of the deconvolved profiles from observations on 2001
August 9 (left) and 2001 August 10 (right). The orbital motion has
been removed assuming the binary parameters found in
Section~\protect\ref{sec:pars}, which allows the individual starspot
tracks across the profiles and the variation in $v \sin i$ to be more
clearly observed. In order to increase the contrast and allow the
starspot features in the trailed spectrum to be more readily seen, a
theoretical profile (see text) has also been subtracted from each LSD
profile before trailing. In this case, starspot features now appear
dark in the plots. Once again, the small gaps in the phase coverage are at
times when arc spectra for the purpose of wavelength calibration were
taken and gaps between individual exposures due to readout time are
not shown.}
\label{fig:theor}
\end{figure}

\section{Roche tomography}
\label{sec:rochey}

Roche tomography (\citealt{rutten94}; \citealt{dhillon01};
\citealt{watson01a}; \citealt{watson03}) is analogous to the Doppler
imaging technique used to map rapidly rotating single stars
(e.g. \citealt{vogt83}). Specifically, Roche tomography is designed to
map the Roche-lobe-filling donor stars in interacting binaries such as
CVs and X-ray binaries.  In Roche tomography, the secondary star is
assumed to be Roche-lobe-filling, locked in synchronous rotation and to
have a circularized orbit. The secondary star is then modelled as a
series of quadrilateral tiles or surface elements, each of which is
assigned a copy of the local (intrinsic) specific intensity
profile. These profiles are then scaled to take into account the
projected area of the surface element, limb darkening and obscuration,
and then Doppler shifted according to the radial velocity of the
surface element at that particular phase. Summing up the contributions
from each element gives the rotationally broadened profile at that
particular orbital phase.

By iteratively varying the strengths of the profile contributed from
each element, the `inverse' of the above procedure can be
performed. The spectroscopic data is fit to a target $\chi^2$, and a
unique map is selected by employing the maximum-entropy {\sc memsys}
algorithm developed by \citet{skilling84}. Unless otherwise stated in
the text, we employ a moving uniform default map, where each element
in the default is set to the average value in the reconstructed map.
Since the maximum-entropy algorithm we have adopted requires that the
input data is positive, we have inverted the absorption-line profiles
and set any negative points to zero. For further details of the Roche
tomography process see the review by \citet{dhillon01}.

Several Doppler imaging studies have used a two-temperature or
filling-factor model (e.g. \citealt{cameron94}) during the image
reconstruction process. These studies work on the assumption that
there will only be two-temperature components across the surface of
the star, those of the immaculate photosphere and those of the cooler
spots. Each image pixel is then effectively made up of a combination
of photosphere and spot intensity. The benefits of using a two-spot
model are that it `thresholds' the maps, preventing the growth of
bright pixels (e.g. \citealt{hatzes92}), and preserves the total spot
areas (\citealt{cameron92}) making quantitative analysis of the maps
more easy.

With Roche tomography we may well expect the secondary star to display
a much more complex topography than isolated stars. In particular, the
surface temperature across a CV secondary will vary greatly not just
due to starspots but also as a result of irradiation, which is
prominent in these systems even in single-line studies
(e.g. \citealt{davey92}; \citealt{davey96}; \citealt{watson03}).  For
this reason, we cannot restrict our Roche tomograms to a
two-temperature reconstruction. Thus Roche tomograms present images of
the distribution of line flux on the secondary star -- effectively
showing the contrast between the various features, blurred by the
action of the regularisation constraint.

\section{Ephemeris}
\label{sec:ephemeris}

A new ephemeris for AE Aqr was determined by cross-correlation with a
template star of spectral type K4V. We only considered the spectral
region containing the strong Mg {\sc ii} triplet absorption for
simplicity. The AE Aqr and K4V template spectra were first
normalised by dividing by a constant, and the continuum was then
subtracted off using a third order polynomial fit. This procedure
ensures that the line strength is preserved along the spectral region
of interest.

The template spectrum was then artificially broadened to account for
the rotational velocity ($v \sin i$) of the secondary star.  The
projected rotational broadening of the secondary star was measured
using an optimal-subtraction technique in which the template spectrum
was broadened, multiplied by a constant and then subtracted from an
orbitally-corrected AE Aqr spectrum. The process was repeated in an
iterative process, adjusting the broadening at each step, until the
scatter in the residual spectrum was minimised.

Through the above process a cross-correlation function (CCF) was
calculated for each AE Aqr spectrum over the wavelength
region covering the strong Mg {\sc ii} triplet absorption. The radial
velocity curve in Fig.~\ref{fig:rvel} was then obtained by fitting a
sinusoid through the CCF peaks to obtain a new zero-point for the
ephemeris of
\begin{equation}
T_0 =\, $HJD$\, 2452131.31345 \pm 0.00007
\end{equation}
with the orbital period fixed at $P=0.41165553$ d (from
\citealt{casares96}). All subsequent analysis has been phased with
respect to this new zero-point. This method of deriving the ephemeris
is relatively insensitive to the use of an ill-matching template, or
incorrect amount of broadening. Indeed, it is more likely to be
dominated by the effects of irradiation, which is well known to
introduce systematic errors in radial velocity studies if not properly
accounted for (e.g. \citealt{davey92}; \citealt{davey96}).  Although
the steeper slope of the radial velocity curve around phase 0.5
suggests the presence of irradiation which is confirmed in our Roche
tomograms (albeit at a low level -- see section{~\ref{sec:maps}), the
ephemeris derived here provides a substantially improved image quality
(both in terms of final reduced $\chi^2$ and artefacts in the
reconstructed map) when compared against that achieved when using the
zero-point of the ephemeris published by \citet{casares96}.

We should note that the work in this section was carried out for the
sole purpose of determining a revised ephemeris in order to improve
the quality of the Roche tomograms in section~\ref{sec:maps}. From
this perspective, no detailed attempt was made to determine the
best-fitting spectral type and accurately measure the rotational
broadening and other parameters from this analysis -- a more
traditional study along these lines will be presented in a future
paper. For completeness, however, we obtained a secondary star radial
velocity amplitude of $K_r$ = 168.4 $\pm$ 0.2 km s$^{-1}$ and a
systemic velocity of $\gamma$ = --59.0 $\pm$ 1.0 km s$^{-1}$ (assuming
the K4V template star HD24916A has a systemic velocity of +6.2$\pm$1.0
km s$^{-1}$, \citealt{king03}). As discussed later in
section~\ref{sec:pars} these measurements will be biased due to, for
example, surface inhomogeneities across the secondary star and the
incorrect treatment of the absorption-line profile shape (which
assumes that the star is spherical). Thus, the parameters derived from
the radial velocity curve have not been used in the subsequent
analysis presented in this paper.

\begin{figure}
\psfig{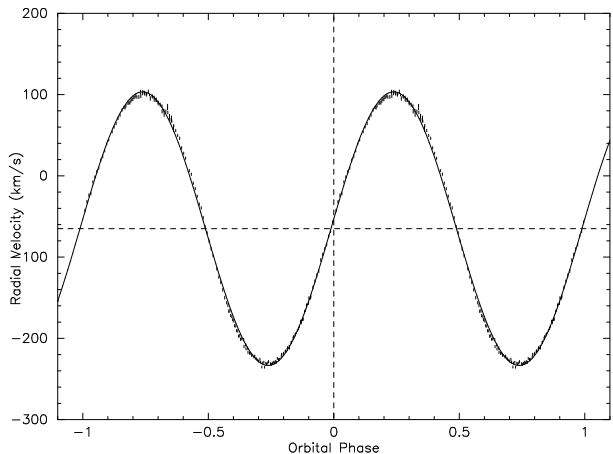}
\caption{The measured secondary star radial-velocity curve with the
best-fitting sinusoidal fit for 2001 August 9 and 10. The data
have been phase folded and then repeated.}
\label{fig:rvel}
\end{figure}

\section{system parameters}
\label{sec:pars}

As discussed in \citet{watson01a}, adopting incorrect system
parameters such as systemic velocity, component masses and orbital
inclination when carrying out a Roche tomography reconstruction
results in spurious artefacts in the final map. These artefacts
normally manifest themselves in the form of bright and dark streaks
and cause a decreased (more negative) entropy to result in the final
map. In other-words, adopting incorrect parameters causes {\em more}
structure to be mapped onto the secondary star. By carrying out
reconstructions for many pairs of component masses (iterating to the
same $\chi ^2$ on each occasion) we can construct an `entropy
landscape' (see Fig.~\ref{fig:eland}), where each grid element
corresponds to the entropy value obtained in the reconstruction for a
particular pair of component masses. One then adopts the set of
parameters that gives the map containing the least structure (the map
of maximum entropy).

For the reconstructions carried out here we have assumed a root-square
limb-darkening law of the form

\begin{equation}
I\left(\mu\right) =
I\left(1\right)\left[1-a\left(1-\mu\right)-b\left(1-\sqrt{\mu}\right)\right]
\end{equation}
where $\mu = \cos \gamma$  ($\gamma$ is the angle between the line of
sight and the emergent flux), and $I(1)$ is the monochromatic specific
intensity at the center of the stellar disc. The limb-darkening
coefficients $a=0.647$ and $b=0.213$ were adopted, which is suitable
for a star of $T_{eff}$ = 4800K, $\log g$ = 4.5 at the central
wavelength of our observations (\citealt{claret98}). This form of the
limb-darkening law has been chosen over the more usual linear
dependency since it has been found that model atmospheres of cool
stars show limb-darkening laws which are largely non-linear (e.g.
\citealt{claret98}).

To determine the binary parameters of AE Aqr, we have constructed a
sequence of entropy landscapes like that described above for a range
of orbital inclinations $i$ and systemic velocities $\gamma$. Then,
for each $i$ and $\gamma$, we picked the maximum entropy value in the
corresponding entropy landscape. From Fig.~\ref{fig:gamma} this method
yields a systemic velocity of $\gamma$ = --63 $\pm$ 1 km s$^{-1}$,
where we have adopted the 1 km s$^{-1}$ error bar from the resolution
of our grid search as a guide rather than a rigorous error
estimate. This is consistent with the systemic velocity of
\citet{casares96} who found $\gamma = -60.9 \pm 2.4$ km s$^{-1}$, and
is in excellent agreement with the value of --63 $\pm$ 3 km s$^{-1}$
found by \citet*{welsh95}. The value obtained using Roche tomography
is slightly more discrepant with the value of $\gamma = -59 \pm 1$ km
s$^{-1}$ obtained in section~\ref{sec:ephemeris} from the radial
velocity curve. We note that the value of the systemic velocity
obtained from the entropy landscape technique is constant over a wide
range of assumed orbital inclinations, as found in \citet{watson03}.

\begin{figure}
\psfig{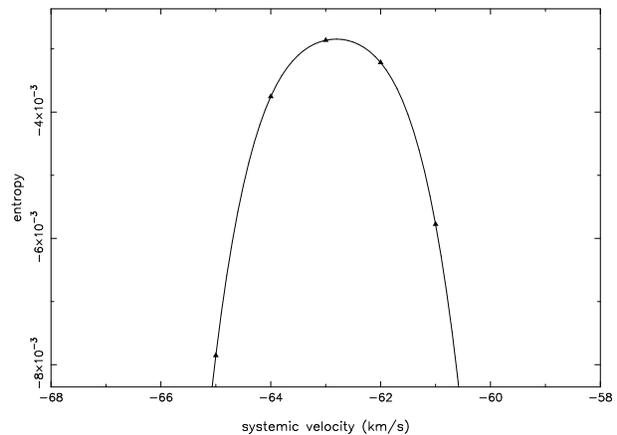}
\caption{Points show the maximum entropy value obtained in each
entropy landscape as a function of systemic velocity, assuming an
orbital inclination of 66$^{\circ}$. (The optimal value for the
systemic velocity is, however, found to be independent of the assumed
inclination). The solid curve shows a fourth order polynomial fit
through these points with the intention to show more clearly where the
peak in entropy occurs with respect to the assumed systemic velocity
in the reconstructions.}
\label{fig:gamma}
\end{figure}

\begin{figure}
\psfig{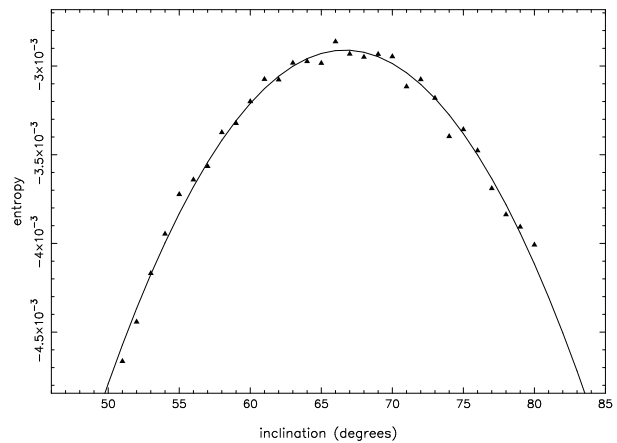}
\caption{Points show the maximum entropy value obtained in each
entropy landscape for different inclinations, assuming a systemic
velocity of $\gamma$ = --63 km s$^{-1}$. The solid curve shows a third
order polynomial fit through these points. The entropy value peaks
at an orbital inclination of $i = 66^{\circ}$.}
\label{fig:incl}
\end{figure}

\begin{figure}
\psfig{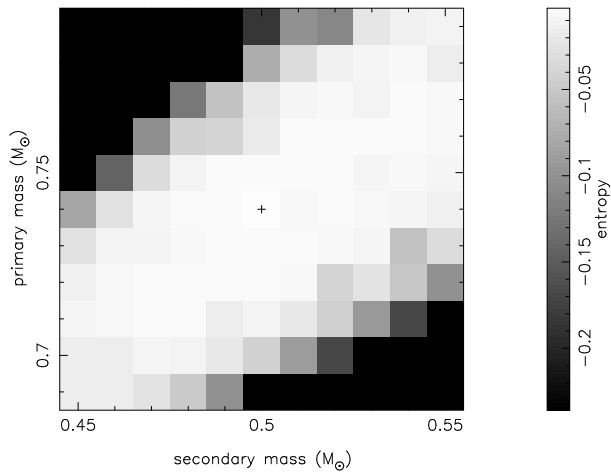}
\caption{The entropy landscape for AE Aqr, assuming an orbital
inclination of $i = 66^{\circ}$ and a systemic velocity of $\gamma =
-63$ km s$^{-1}$. Both nights data were combined when carrying out the
reconstructions. Dark regions indicate masses for which no
acceptable solution could be found.  The cross marks the point of
maximum entropy, corresponding to component masses of $M_1$ = 0.74
M$_{\odot}$ and $M_2$ = 0.50 M$_{\odot}$.}
\label{fig:eland}
\end{figure}

Fig.~\ref{fig:incl} shows the maximum-entropy value as a function of
inclination obtained in entropy landscapes assuming the systemic
velocity of $\gamma = -63$ km s$^{-1}$ derived above. This indicates a
distinct increase in the amount of structure present in the maps
either side of the best fitting inclination at $i=66^{\circ}$, which
we have adopted as the optimal value for the orbital inclination of AE
Aqr. It is comforting to see that this lies below the $i<70^{\circ}$
limit inferred from the lack of eclipses (\citealt*{chanan76}), though
it is still relatively high compared to previous
estimates. \citet{casares96} determined $i=58^{\circ} \pm 6^{\circ}$
by modelling the phase dependent variation of the rotational
broadening in their intermediate resolution spectra, while
\citet{welsh95} constrained $i$ to $55^{\circ} \pm 7^{\circ}$ in their
absorption line study of AE Aqr.

\citet{shahbaz98} noted, however, that the methods used by
\citet{welsh95} and \citet{casares96} to measure the orbital
inclination of AE Aqr are prone to systematic errors and result in a
shift to lower values for the inclination. This systematic error
results from the incorrect treatment of the absorption-line profile
shape which is often approximated by using the Gray profile
\citep{gray92} for spherical stars. Since the determination of the
radial velocity amplitude and rotational broadening of the secondary
star is also sensitive to the correct treatment of the line-profile --
use of the Gray profile or failure to account for surface
inhomogeneities on the secondary star can lead to systematic errors in
the derived mass ratio as well. Roche tomography, however, predicts
the correct broadening function for a Roche-lobe filling star and, in
addition, takes into full account the line flux distribution across
the star which can lead to systematic errors in these more
conventional studies. Further support for the higher inclination found
in this work comes from a recent preliminary analysis of high
spectral-resolution data by \citet{echevarria05}.  These authors have
also found evidence for a high ($\sim$70$^{\circ}$) inclination from
studying the phase-dependent variation of the rotational broadening.

The entropy landscape for AE Aqr assuming an orbital inclination of
$i$ = 66$^{\circ}$ and a systemic velocity of $\gamma$ = --63 km
s$^{-1}$ is shown in Fig.~\ref{fig:eland}. From this we derive a
secondary star mass of $M_2$ = 0.50 M$_{\odot}$ and a primary mass of
$M_1$ = 0.74 M$_{\odot}$. These are in good agreement with the most
recent previous estimates in the literature, including $M_1$ = 0.79
$\pm$ 0.16 M$_{\odot}$ and $M_2$ = 0.50 $\pm$ 0.10 M$_{\odot}$
(\citealt{casares96}), as well as $M_1$ = 0.89 $\pm$ 0.23 M$_{\odot}$
and $M_2$ = 0.57 $\pm$ 0.15 M$_{\odot}$ (\citealt{welsh95}).

In all cases we have carried out reconstructions to $\chi^2$ = 0.8,
though we have explored the effects of carrying out the
reconstructions to different $\chi^2$ values.  For reconstructions to lower
$\chi^2$ values the entropy landscapes break up into noise, whereas when
iterating to higher $\chi^2$ values the entropy landscapes favour higher
inclinations (which are unphysical given the lack of eclipses) --
eventually losing all sensitivity to the assumed inclination. In all
cases, however, the derived component masses for any given orbital
inclination all agree to better than 0.02M$_{\odot}$ for all
reasonable selections of target $\chi^2$. We are confident that we are
neither over-fitting nor under-fitting the data. The maps rapidly
break into noise for $\chi^2$ values less than 0.8, leading to noisy
entropy landscapes. Maps fit to greater $\chi^2$ values become `washed out'
with features becoming less distinct as more pixels are assigned the
default value in the absence of proper data constraints.

It should also be noted that we have not assigned errors to our system
parameter estimates. Technically, this could be done using a
Monte-Carlo style technique and applying the bootstrap resampling
method (\citealt{efron79}; \citealt{efron93}) to generate synthetic
datasets drawn from the same parent population as the observed dataset
(see also \citealt{watson01a}).  Unfortunately, this would mean
repeating the analysis carried out in this paper for several hundred
bootstrapped datasets which would require several months of
computational time. It is, therefore, unfeasible for us to assign
strict errors bars to our derived binary parameters. Given the quality
of the data (high time and spectral-resolution) and the full treatment
of the data by Roche tomography (which takes into account the
roche-lobe shape and any surface inhomogeneities), we believe that the
system parameters derived in this paper for AE Aqr are the most
accurate to date.

\section{surface maps}
\label{sec:maps}

Using the binary parameters derived in Section~\ref{sec:pars} and
combining both nights datasets, we have constructed a Roche tomogram
of the secondary in AE Aqr, and this is displayed in
Fig.~\ref{fig:map1}. The corresponding fit to the datasets obtained on
both nights are displayed in Fig.~\ref{fig:fits}.  Dark spots are
evident in the Roche tomogram. The most noticeable feature
is the large starspot feature which lies near (but not centred on) the
polar regions at a latitude of $\sim$65$^{\circ}$ and is skewed
towards the trailing hemisphere ($\phi$=0.25) of the star. Such
high-latitude and polar spots are common in Doppler images of other
rapidly-rotating stars and are discussed in more detail in
Section~\ref{sec:discussion}.

The other prominent dark region in the tomogram is best seen at
$\phi$=0.5. This lies at the $L_1$ point on the secondary star, and
could possibly be due to irradiation from the accretion regions which
would also appear dark in the tomograms since the weak metal lines
become ionised.  Although irradiation of the donor star in AE Aqr has
not been seen previously, it's location at the $L_1$ point is
consistent with that seen in other CVs
(e.g. \citealt{watson03}). Furthermore, in high spectral-resolution
observations taken at the Magellan 6.5-m telescope during July 2004
(to be presented in a later paper), the preliminary LSD profiles
around $\phi$=0.5 are almost identical to those presented here. This
is indicative of a fixed feature around the $L_1$ point which is more
likely due to the presence of irradiation rather than the existence of
a long-lived (several year) spot feature. On the other-hand, we
currently do not have a knowledge of the spot life-times for this
class of star and the modelling of the mass transfer history of AM Her
by \citet{hessman00} suggests that the spottedness of the $L_1$ point
on the donor stars in CVs may be untypical compared to isolated
stars. Despite these caveats, however, irradiation is still our
preferred interpretation of the feature located at the $L_1$ point.

In addition to these prominent features, a number of other smaller
spots can be seen at lower latitudes. Unfortunately many of these
lower latitude spots are heavily smeared across the equator since the
latitudinal resolution of the imaging technique decreases towards the
equator (this is because the path of a starspot trail in the time series
shows the least variation as a function of latitude near the equatorial
regions). This problem is further compounded by the lower
signal-to-noise of our LSD profiles relative to many other Doppler
imaging studies of isolated stars (a result of the faintness of AE
Aqr), and the target's relatively high orbital inclination, which
results in a degeneracy between which hemisphere a feature is located
on (e.g. \citealt{watson01a}). Therefore, it is difficult to tell
whether the lower latitude spots are located within a particular
latitude band or not.

To check whether or not the features seen in the Roche tomogram are
real, or spurious artefacts due to noise, we have carried out two
further Roche tomography constructions using only the odd-numbered
spectra for one and the even-numbered spectra for the other (see
Fig.~\ref{fig:map2}).  Despite the reduction in phase-sampling, both
of these Roche tomograms show the same features as seen in
Fig.~\ref{fig:map1} and therefore we are confident that these features
are real. Furthermore, when we construct separate maps for each nights
data (not shown here for reasons outlined below), we again find that
that features are reconstructed at similar locations in the maps where
there is sufficient phase overlap in the observations. This gives us
further confidence in the reality of the reconstructed features.

Although the maps reconstructed on individual nights show similar
features, these features do differ in detail. Unfortunately, this is
most likely a result of the relatively poor signal-to-noise of our
observations combined with the introduction of phase gaps when
reconstructing each dataset separately -- further compounded by the
deteriorating seeing conditions on night 2. We feel, therefore, that it
would be unwise to speculate as to whether any of the detailed
differences between the individual nights reconstructions are due to
the evolution of spots other than to mention that there does not
appear to be any significant movement of spot features. Future higher
signal-to-noise observations will be better placed to resolve spot
evolution.

In order to make a more quantitative estimate of the spot parameters
on AE Aqr, we have examined the intensity distribution of the pixels
in the Roche tomogram. By looking for a bimodal distribution in pixel
intensities, an estimate of the spot properties can be obtained by
labelling low-intensity pixels as spotted regions, and high-intensity
pixels as immaculate photosphere. First, we discarded all pixels on
the southern hemisphere as this hemisphere is least visible and as
such will mainly contain pixels assigned with the default intensity
which, in this case, will be the average intensity of the
map. Inclusion of these pixels would `wash-out' any underlying bimodal
distribution in pixel intensity. The corresponding histogram of pixel
values is shown in Fig.~\ref{fig:hist} and, as expected, this shows a
group of pixels of lower intensity which have been labelled as
starspots (see caption of Fig.~\ref{fig:hist} for details).

Although far from an ideal means of classifying what regions are
spotted and what regions contain immaculate photosphere, this is the
best classification scheme given our inability to apply a
two-temperature model as discussed in Section~\ref{sec:rochey}. Based
on this scheme, Fig.~\ref{fig:scover} shows the distribution of
starspots as a function of latitude scaled by either the total
surface area of the star, or by the surface area of the
latitude strip over which the spot coverage was calculated. This
confirms our visual inspection of the tomogram and that the density of
spots increases towards the pole, with a grouping of spots centred
around a latitude of +$65^{\circ}$ corresponding to the large spot
seen near the north pole.

Inspection of Fig.~\ref{fig:scover} also shows an apparent depletion
of starspots near +$40^{\circ}$, with an increase in starspot coverage
again towards lower latitudes, with a further increase in spot
coverage below +$10^{\circ}$. This latter increase in spots below a
latitude of $\sim10^{\circ}$ is probably due in part to the presence
of irradiation near the $L_1$ point. Nonetheless, the irradiated
region is not large enough to fully explain the increase in
low-latitude pixels being flagged as spots. Therefore, this
indicates the presence of a lower latitude site of magnetic activity
on AE Aqr which future higher signal-to-noise observations will help
resolve.

In total, we estimate that 18 per cent of the northern hemisphere on
AE Aqr is spotted. We stress that this number should be regarded with
caution given our scheme for classifying pixels containing spots, but
that such a figure for the spot filling-factor is not at all
unreasonable and, if anything, is likely to be an under-estimate of
the true spot filling factor. TiO studies of rapidly rotating stars by
\citet{oneal98} have revealed spot filling factors as large as 50 per
cent in the case of II Peg.  Generally, such studies result in much
larger filling factors than are deduced from Doppler images which
typically return spot-filling factors from $\sim$4 per cent (e.g.
Speedy Mic, 4.5 per cent, \citetalias{barnes01a}) to $\sim$12 per cent
(e.g.  He 699, 11.9 per cent, \citetalias{barnes01b}).

A prime example of the underestimation of the global spot coverage by
Doppler imaging methods was highlighted by \citet{barnes04}, whose
Doppler images of the primary star in the contact binary AE Phe returned
a spot filling factor of $\sim$5 per cent. \citet{barnes04}, however,
estimated that the actual spot coverage could be 27 per cent in order
to lower the apparent temperature of the primary component to the
level seen in their observations. This discrepancy between Doppler
imaging and, for example, TiO studies is probably caused by the
failure of Doppler imaging to resolve small starspots, which will
certainly be the case for the lower signal-to-noise dataset presented
in this paper. The poorer latitudinal resolution near the equator will
also cause low-latitude spots to be heavily smeared over a larger
surface area, reducing their contrast with the immaculate photosphere
and increasing the probability of the non-detection of starspots at
these lower latitudes.  Finally, \citet{webb02} found a spot filling
factor of 22 per cent from a TiO study of the CV SS Cyg, similar to
our estimate for AE Aqr.

\begin{figure}
\psfig{figure=aeaqr_map.ps,width=8.0cm,angle=-90.}
\caption{Roche tomogram of AE Aqr. Dark greyscales depict regions of
suppressed absorption and are indicative of starspots or irradiated
zones. The system is plotted as the observer would see it at an
orbital inclination of 66$^{\circ}$, except for the central panel
which shows the system as viewed from above the North pole. The
orbital phase (with respect to the ephemeris of
Section~\protect\ref{sec:ephemeris}) is indicated above each panel.
For clarity, the Roche tomograms are shown excluding limb-darkening.}
\label{fig:map1}
\end{figure}

\begin{figure*}
\begin{tabular}{c}
\psfig{figure=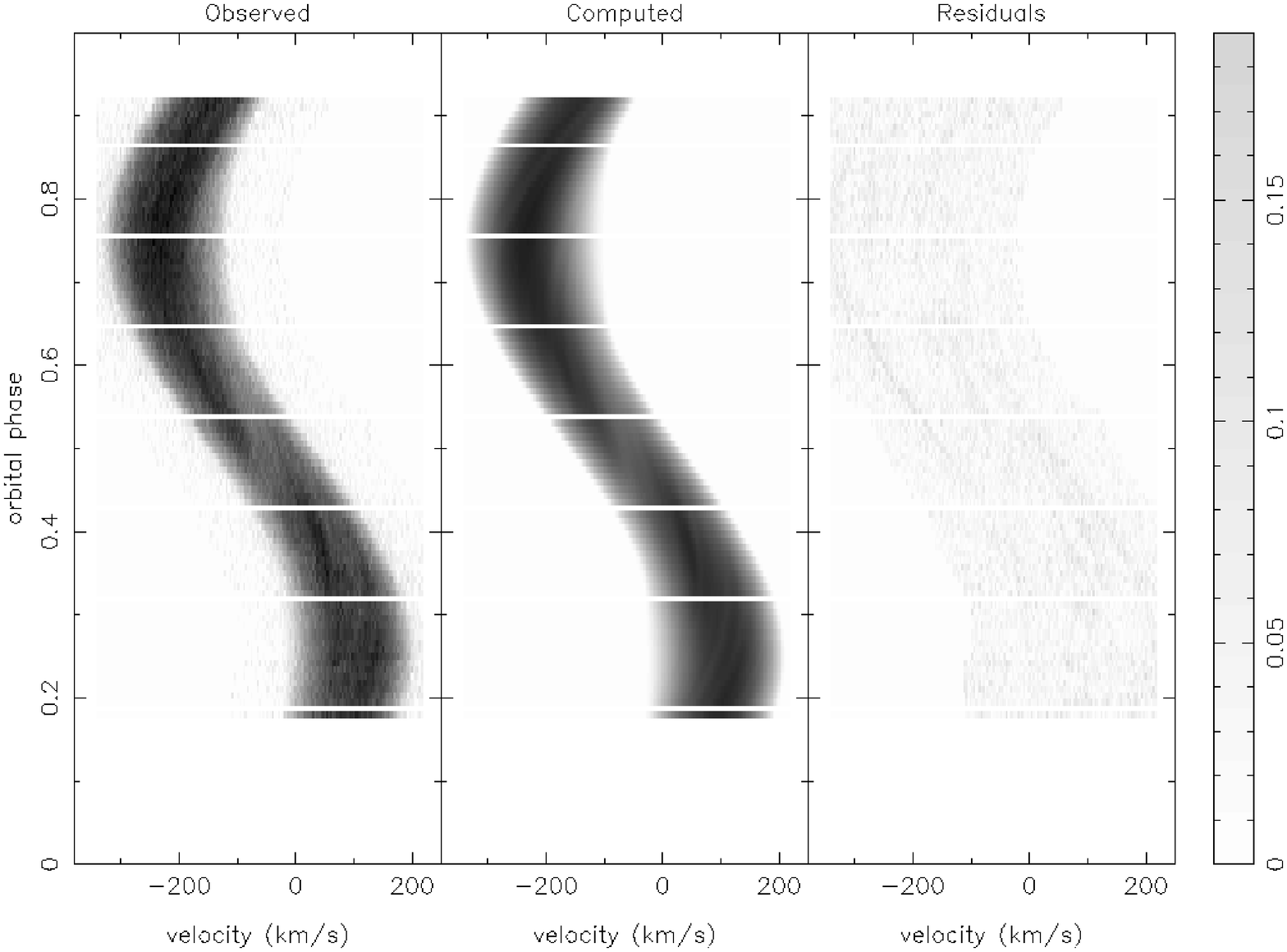,width=14.5cm,angle=-90.} \\
\vspace{0.4cm} \\
\psfig{figure=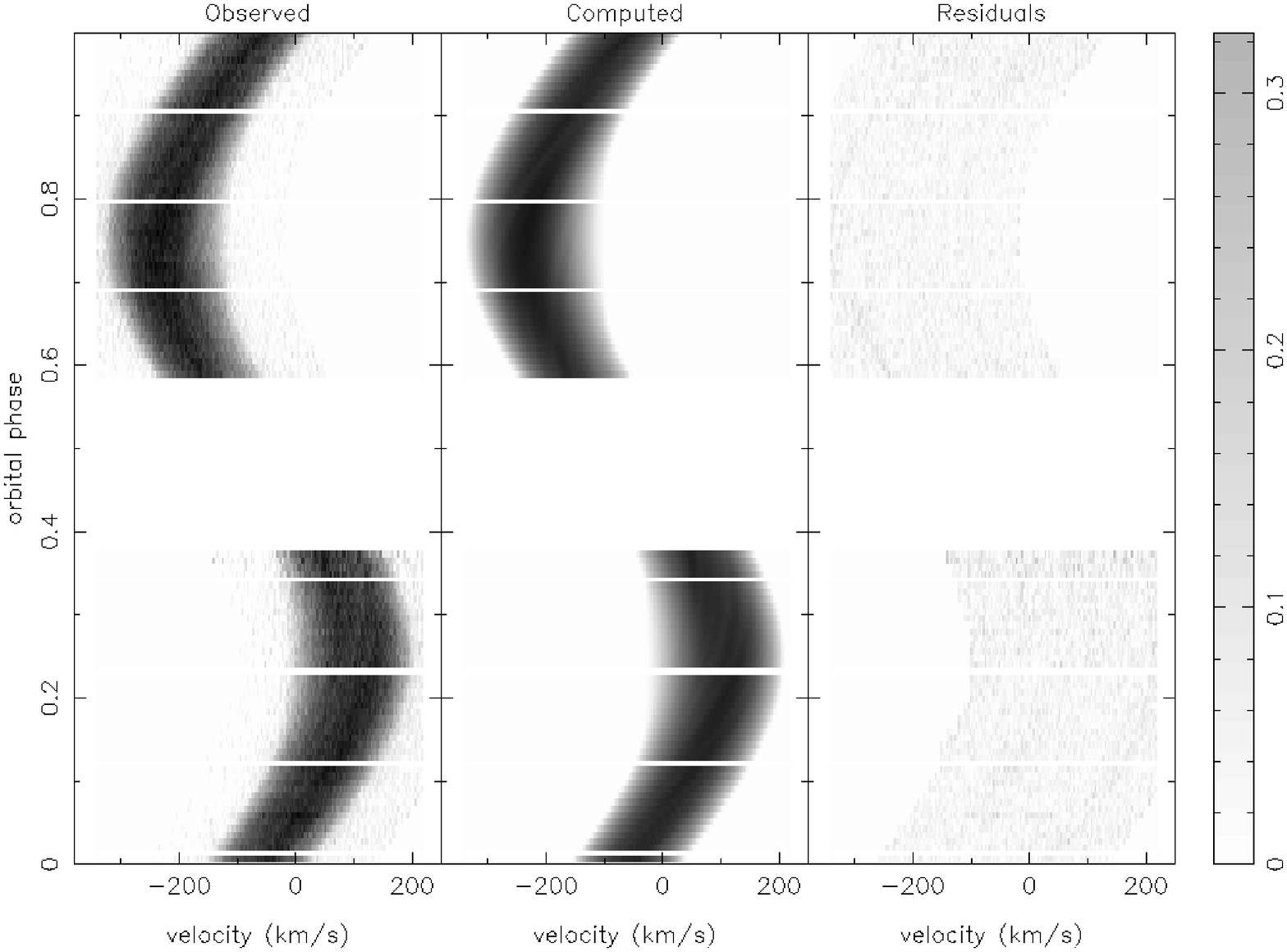,width=14.5cm,angle=-90.} \\
\end{tabular}
\caption{Roche tomography fits to the LSD trailed spectra showing the
observed trails, computed trails from the Roche tomography
reconstruction and the O--C residuals. The greyscale wedge (right)
indicates the magnitude of the residuals as a fraction of the average
line profile depth. The largest residuals result from spikes in the
deconvolved continuum (which do not adversely affect the Roche
tomography fits to the profiles themselves - see
Figs~\protect\ref{fig:profs1} \& \protect\ref{fig:profs2}). The
residuals are greatest for the second nights observations due to the
deterioration in the seeing at the end of the night. Top: observations
from 2001 August 9. Bottom: observations from 2001 August 10.}
\label{fig:fits}
\end{figure*}

\begin{figure}
\psfig{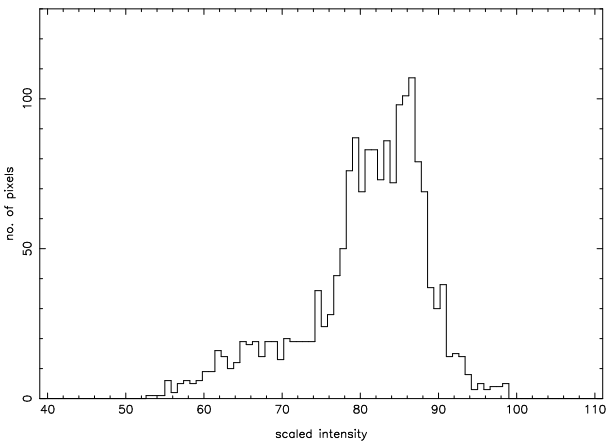}
\caption{Histogram of the pixel intensities from the Roche tomogram of
AE Aqr after pixels on the southern hemisphere (latitude $<$
0$^{\circ}$) were discarded. The brightest pixel in the map is
assigned an intensity of 100 and other pixel intensities are scaled
linearly against this. For the purposes of estimating the global spot
properties of AE Aqr, the immaculate photosphere has been defined as
pixels with intensities of 74 or greater, and the spotted photosphere
those regions where pixel intensities are less than 74.}
\label{fig:hist}
\end{figure}

\begin{figure*}
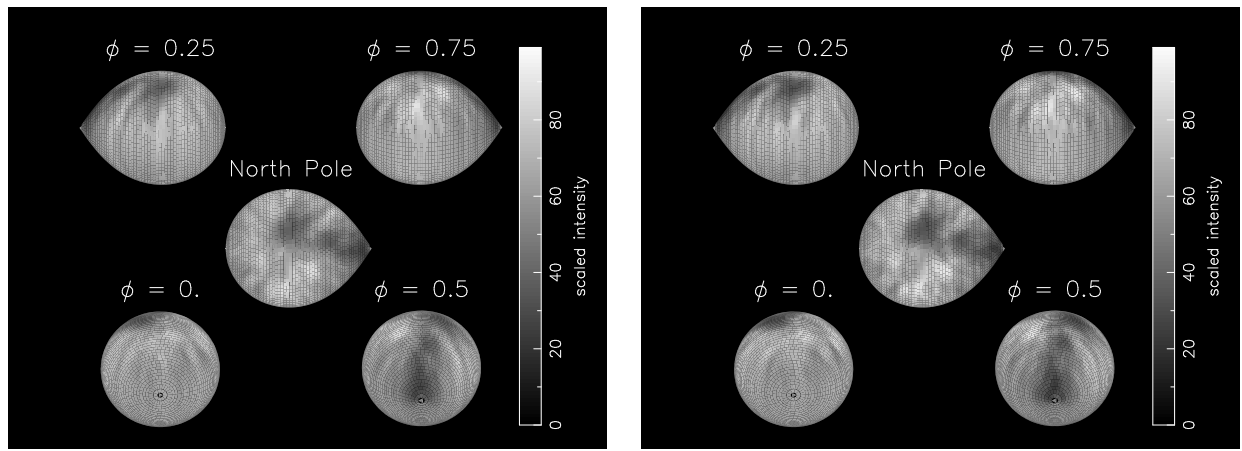

\begin{tabular}{ll}
\psfig{figure=aeaqr_odd.ps,width=8.0cm,angle=-90.} &
\psfig{figure=aeaqr_even.ps,width=8.0cm,angle=-90.} \\
\end{tabular}
\caption{As for Fig.~\protect\ref{fig:map1}, but reconstructions for
odd-numbered spectra (left) and even-numbered spectra (right).}
\label{fig:map2}
\end{figure*}

\begin{figure*}
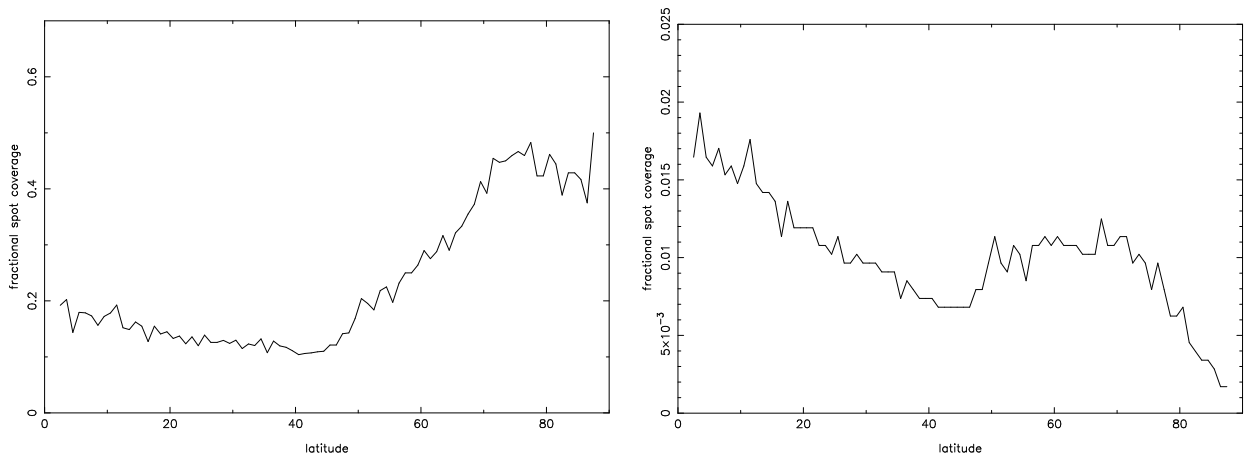

\begin{tabular}{ll}
\psfig{figure=spotcoverage.ps,width=8.0cm,angle=-90.} &
\psfig{figure=spotcoverage2.ps,width=8.0cm,angle=-90.} \\
\end{tabular}
\caption{Plots showing the spot coverage as a function of
latitude. Left: spot coverage as a function of latitude expressed in
terms of the surface area at that latitude. Right: spot coverage  as a
function of latitude normalised by the total surface area of the
northern hemisphere. The curves appear noisy in comparison to similar
plots elsewhere in the literature as the elements that make up our
stellar grid are not aligned along constant latitude strips in order
to ensure that the inner-Lagrangian point is tiled
adequately. Therefore a certain degree of interpolation in latitude
between adjacent tiles is required to calculate the contribution of
each tile to that latitude.}
\label{fig:scover}
\end{figure*}

\section{Discussion}
\label{sec:discussion}

The secondary in the cataclysmic variable binary AE Aqr is a highly
active star as evinced by the heavily spotted Roche tomogram presented
in this work.  This is unsurprising, given the rapid (0.41-d) rotation
period of AE Aqr.  In keeping with the abundant observations of
high-latitude or polar spots on rapidly rotating stars (see
\citealt{schrijver01} and references therein), AE Aqr also exhibits a
large spot at high-latitude. This spot is, however, not truly polar as
seen in many Doppler images as it is centred at a latitude of around
+$65^{\circ}$. This is still in striking contrast to our Sun, where
starspots seldom appear beyond $\pm30^{\circ}$ from the equator
(e.g. \citealt{pulkkinen99}). After much debate about the reality of
the polar spots revealed by Doppler imaging (see \citealt{hatzes96},
\citealt{unruh97},  \citealt{bruls98}), the reality of these features
now appear to be established within the field. \citet{schussler92}
have suggested that such high latitude spots are the cause of the
strong Coriolis forces in these rapidly rotating stars which drive the
magnetic flux tubes towards the polar regions. Alternatively, such
spots can form at lower latitudes and then migrate polewards - such a
poleward migration of mid-latitude spots has been reported for the RS
CVn binary HR1099 (V711 Tauri -- \citealt{vogt99},
\citealt{strassmeier00}). Future observations will be needed in order
to establish whether the starspots on AE Aqr also undergo a poleward
motion.

In addition, AE Aqr also appears to show starspots at latitudes below
$\sim35^{\circ}$. As described in \citet{granzer00}, such bimodal spot
distributions (the simultaneous appearance of spots at high and
intermediate/low latitudes) have also been reported for the zero-age
main-sequence (ZAMS) rapid rotator AB Dor (e.g. \citealt{unruh95b}),
the Pleiades-age main-sequence star LQ Hya (\citealt{rice98},
\citealt{donati99b}), the $\alpha$ Per G-dwarves He669 and He520
(\citealt{barnes98}), and the young G-dwarf EK Dra
(\citealt{strassmeier98}) amongst others.

In their paper studying the distribution of starspots on cool stars,
\citet{granzer00} found that low-mass stars ($\sim 0.4$M$_{\odot}$)
with rotation rates more than 4 times solar showed spot emergence at
latitudes between $\sim60-70^{\circ}$. It is perhaps worth noting that
this is comparable with AE Aqr which, at 0.5M$_{\odot}$, shows spots
emerging at $\sim65^{\circ}$. Unfortunately, since AE Aqr has already
evolved some way off the ZAMS, comparison to the results from
modelling young stars by \citet{granzer00} may make any conclusion drawn
from such a comparison somewhat uncertain.

It is also interesting to compare the images of AE Aqr to those of the
late pre-main sequence or early ZAMS isolated star Speedy Mic
\citepalias{barnes01a}. Excluding their ages and the effect of tidal forces,
Speedy Mic (of spectral-type K3V and rotation period 0.38-d) otherwise
has similar stellar parameters to AE Aqr ($P_{rot}$ = 0.41-d,
K3-5V/IV).  Like AE Aqr, Speedy Mic does not show any truly polar spot
but does show a concentration of starspots around 60--70$^{\circ}$ at
a similar latitude to the large spot group observed on AE Aqr at
65$^{\circ}$. Speedy Mic also exhibits a bimodal distribution of
starspots, with a similar concentration of spots at lower latitudes to
that observed for AE Aqr. Furthermore, Barnes et al. (2001a) also describe
a relative paucity of spots at a latitude of 30--40$^{\circ}$, the
upper limit of which coincides with the depletion of spots around
40$^{\circ}$ seen on AE Aqr.  Such depletion of spots at intermediate
latitudes is also observed in the K2 dwarf LQ Hya (\citealt{rice98},
\citealt{donati99b}) and the K5 dwarf BD +55$^{\circ}$ 4409 (LO Peg --
\citealt*{lister99}). More observations of late-type stars are
required in order to determine whether this feature is common to all
K-type and later stars.

In conclusion, we have unambiguously imaged, for the first time, large
starspots on the secondary star in AE Aqr, the distribution of which
is similar to other rapidly rotating low-mass stars. We estimate a
spot coverage of approximately 18 per cent, indicating a high-degree of
magnetic activity which is no doubt due to an efficient dynamo driven
by rapid rotation. Such a high-level of activity on AE Aqr provides
evidence that magnetic-braking in long-period CVs could play a significant
role in their evolution towards the period gap.

\section{Future opportunities}
\label{opportunities}

Given that this is the first time that starspots have been imaged on
the secondary stars in CVs, we thought it prudent to discuss the
future avenues that such studies could explore.

\subsection{Spot properties on CV secondaries}

As briefly mentioned in Section~\ref{sec:intro}, a knowledge of
the distribution of starspots on Roche-lobe filling stars
can provide tests of our understanding of stellar dynamos.
For example, the models of \cite{holzwarth03} show that the tidal
effects in a binary due to the presence of a close companion
can lead to the formation of clusters of flux tubes (and therefore
starspots) at preferred longitudes. Furthermore, these effects
are predicted to become more pronounced for shorter period systems
where the tidal forces are strongest. Since CVs contain some of
the most rapidly rotating stars known, they provide excellent
candidates in which to study the effects of tidal forces on
magnetic flux tube emergence. In addition, since the magnitude
of the tidal effects will change with changing binary parameters,
so will the predicted starspot distributions. Therefore it
is highly desirable to obtain a set of high-resolution maps of
a number of different CVs as each individual binary system studied
should, in principle, provide an independent test of such models.

Studies should also be carried out to investigate how the starspot
coverage varies with time in an attempt to deduce whether these stars
display magnetic activity cycles similar to the 11-year solar
cycle. An understanding of these cycles is crucial to the success of
any stellar dynamo theory. As outlined in Section~\ref{sec:intro},
magnetic activity cycles have also been invoked to explain
quasi-cyclical variations in orbital periods, quiescent magnitudes
and outburst intervals, durations and shapes, with cycles on timescales
as short as $\sim$ 3 years \citep*{ak01}. No {\em direct}
evidence for magnetic activity cycles has so far been found
for these systems and therefore long-term monitoring of a
selection of CVs is actively encouraged. To this end we have
initiated a long-term monitoring campaign of AE Aqr, the first
results of which are to be published in a future paper.

\subsection{Differential rotation of tidally distorted stars}
\label{sec:diffrot}

Differential rotation is thought to play a crucial role in
amplifying and transforming the initial poloidal magnetic field
into toroidal field through dynamo processes (the so-called $\Omega$
effect). Thus measurements of differential rotation rates aim to
evaluate the influence of stellar parameters on the surface shear
rates and hence test stellar dynamo theories. Theoretical work
regarding differential rotation in tidally distorted stars by
\cite{scharlemann82} suggests that it should be weakened,
but not entirely suppressed, in these objects. Some two decades
later, however, observational evidence for the strength of
differential rotation in tidally distorted stars is still confused.
\cite{petit04} find evidence for weak differential rotation
in the RS CVn system HR 1099. Images of the contact binary AE Phe
\citep{barnes04}, on the other-hand, shows rapid spot evolution
and motions. Evidently more studies of tidally distorted systems
are required.

By acquiring two images of a CV separated by several days, one can
cross-correlate strips of constant latitude, allowing the
sense and magnitude of the differential rotation to be
measured. Furthermore, in the event that surface shear exists,
\cite{scharlemann82} note that, on average, the stellar
envelope should still be forced to co-rotate. Roche tomography
studies could also be used to deduce the co-rotation latitude
at which the envelope is tidally-locked to the orbital
period.

Although fainter than RS CVn systems, CVs still offer an excellent
laboratory to study differential rotation in tidally distorted
stars. For instance, RS CVn's have orbital periods that typically span
several days, meaning that long observations covering several nights
are required to establish a complete picture of the system. During
this time individual starspots may have appeared, evolved into
different shapes or disappeared entirely. Therefore a single image of
an RS CVn system can often not be regarded as a single `snapshot' of
the spot morphology -- making the detection and accurate determination
of differential rotation rates more complicated. CVs, however, have
shorter orbital periods (typically $<$ 10 hours) and therefore a single
Roche tomogram of a CV should be obtainable in one night for the majority of
cases -- making differential rotation studies more accessible.

\subsection{Impact of activity on accretion dynamics}

Magnetic activity on the secondary stars in CVs is also thought to
play a role in the accretion dynamics of interacting binaries over
short timescales. \cite{livio94} have suggested that the low states
observed in some interacting binaries can be explained by the passage
of starspots across the inner-Lagrangian ($L_1$) point (see also
\citealt{king98}). This temporarily quenches the flow of material
through the $L_1$ point, thereby reducing the accretion rate. Indeed,
\cite{hessman00} were able to model the long-term brightness
variations of the cataclysmic variable AM Her as modulations caused by
starspots and conclude that either these systems have a large spot
covering fraction ($\sim$ 0.5) or that the $L_1$ point is unusually
spotty. More recently, \cite{demartino02} also attributed the rapid
variation of accretion in Am Her to starspots. Such mass transfer
variations can also have a marked effect on the response of the
accretion disc in dwarf novae (\citealt{schreiber00}).

There are a number of ways that these ideas can be tested with Roche
tomography. First, the observation of a starspot crossing the $L_1$
point in combination with a simultaneous drop in system brightness
(high-resolution Roche tomography datasets should normally be
accompanied by simultaneous photometry) would provide evidence
supporting these models. Although such an observation may seem
fortuitous, given the large spot coverages seen on rapidly rotating
stars (as much as 50 per cent, \citealt{oneal96}) and, in particular,
given the high spot coverages seen in the Roche tomograms of AE Aqr,
there is a high probability of witnessing such an event. The main
problem with this approach would be assessing whether a feature at the
$L_1$ point was truly a spot or caused by irradiation. One suitable
target is the bright Z Cam type dwarf-nova V426 Oph
(\citealt{hellier90}) which exhibits drops from standstill which may
be due to a decrease in mass transfer.  (In recent high-resolution
Magellan observations starspot features were found in the deconvolved
line profiles of V426 Oph and a Roche tomogram of V426 Oph will be
published in a later paper).

Second, even if an event is not seen directly, given a detailed
knowledge of the starspot distributions and the sense and
magnitude of their motions of the starspots due to
differential rotation (Section~\ref{sec:diffrot}), it should be
possible to predict at what times a starspot will migrate
across the $L_1$ point. Again, an observed dimming at the
predicted times would provide strong evidence in favour
of such models. Furthermore, \cite{king98} note that the
most rapid transitions to low states occur over the course
of a day. The length scale of the mass overflow region
near the $L_1$ point is (e.g. \citealt{pringle85})
\begin{equation}
w \simeq \frac{Pc_s}{2 \pi} \simeq 3 \times 10^8 P_{hr}~{\rm cm},
\end{equation}
where $c_s$ is the sound speed of the gas and $P_{hr}$ is the orbital
period $P$ expressed in hours. In order to cause a transition to a
low-state over one day, a spot boundary would have to move across the
$L_1$ point at a velocity of $\sim$ 3 $\times$ 10$^3 P_{hr}$ cm
s$^{-1}$. By using starspots as surface flow tracers it will be
possible to determine whether the differential rotation rates in CVs
are commensurate with these velocities.

Finally, in the model for AM Her proposed by \cite{hessman00}, they
infer a power-law for the distribution of spot sizes. Using Roche
tomography, one can determine both the numbers and sizes of the
starspots on CV secondary stars, allowing the measurement of the
power-law index for the upper end of the spot size distribution
(e.g. \citealt{cameron01}). This would provide a powerful consistency
test for such models.

In conclusion, not only are Roche tomography studies important for our
further understanding of the diverse behaviour of CVs but they also
promise to provide further insights into stellar dynamos. In order to
provide the best constraints on spot sizes and distributions, and to
extend surveys to more (and almost certainly fainter) CVs,
observations using echelle spectrographs on 6 to 8-m class telescopes
will be required.

\section*{\sc Acknowledgements}

CAW was initially employed on PPARC grant PPA/G/S/2000/00598 and is
now supported by a PPARC Postdoctoral Fellowship. TS acknowledges
support from the Spanish Ministry of Science and Technology  under the
programme Ram\'{o}n y Cajal. The authors acknowledge the use of the
computational facilities at Sheffield provided by the Starlink
Project, which is run by CCLRC on behalf of PPARC.

\bibliographystyle{mn2e}
\bibliography{abbrev,refs}

\begin{thebibliography}{}

\bibitem[\protect\citeauthoryear{Ak, Ozkan \& Mattei}{Ak et~al.}{2001}]{ak01}
Ak T.,  Ozkan M.~T.,    Mattei J.~A.,  2001, A\&A, 369, 882

\bibitem[\protect\citeauthoryear{Andronov, Pinsonneault \& Sills}{Andronov
  et~al.}{2003}]{andronov03}
Andronov N.,  Pinsonneault M.,    Sills A.,  2003, ApJ, 582, 358

\bibitem[\protect\citeauthoryear{Applegate}{Applegate}{1992}]{applegate92}
Applegate J.~H.,  1992, ApJ, 385, 621

\bibitem[\protect\citeauthoryear{Barnes}{Barnes}{1999}]{barnes99}
Barnes J.~R.,  1999, PhD thesis, University of St. Andrews

\bibitem[\protect\citeauthoryear{Barnes \& Collier~Cameron}{Barnes \&
  Collier~Cameron}{2001}]{barnes01}
Barnes J.~R.,  Collier~Cameron A.,  2001, MNRAS, 326, 950

\bibitem[\protect\citeauthoryear{Barnes, Collier~Cameron, James \&
  Donati}{Barnes et~al.}{2000}]{barnes00}
Barnes J.~R.,  Collier~Cameron A.,  James D.,    Donati J.-F.,  2000, MNRAS,
  314, 162

\bibitem[\protect\citeauthoryear{Barnes, Collier~Cameron, James \&
  Donati}{Barnes et~al.}{001a}]{barnes01a}
Barnes J.~R.,  Collier~Cameron A.,  James D.,    Donati J.-F.,  2001a, MNRAS,
  324, 231

\bibitem[\protect\citeauthoryear{Barnes, Collier~Cameron, James \&
  Steeghs}{Barnes et~al.}{001b}]{barnes01b}
Barnes J.~R.,  Collier~Cameron A.,  James D.~J.,    Steeghs D.,  2001b, MNRAS,
  326, 1057

\bibitem[\protect\citeauthoryear{{Barnes}, {Collier Cameron}, {Unruh}, {Donati}
  \& {Hussain}}{{Barnes} et~al.}{1998}]{barnes98}
{Barnes} J.~R.,  {Collier Cameron} A.,  {Unruh} Y.~C.,  {Donati} J.~F.,
  {Hussain} G. A.~J.,  1998, MNRAS, 299, 904

\bibitem[\protect\citeauthoryear{Barnes, Lister, Hilditch \&
  Collier~Cameron}{Barnes et~al.}{2004}]{barnes04}
Barnes J.~R.,  Lister T.,  Hilditch R.,    Collier~Cameron A.,  2004, MNRAS,
  348, 1321

\bibitem[\protect\citeauthoryear{Beskrovnaya, Ikhsanov, Bruch \&
  Shakhovskoy}{Beskrovnaya et~al.}{1996}]{beskrovnaya96}
Beskrovnaya N.~G.,  Ikhsanov N.~R.,  Bruch A.,    Shakhovskoy N.~M.,  1996,
  A\&A, 307, 840

\bibitem[\protect\citeauthoryear{Bruch}{Bruch}{1991}]{bruch91}
Bruch A.,  1991, A\&A, 251, 59

\bibitem[\protect\citeauthoryear{Bruls, Solanki \& Sch{\"{u}}ssler}{Bruls
  et~al.}{1998}]{bruls98}
Bruls J. H. M.~J.,  Solanki S.~K.,    Sch{\"{u}}ssler M.,  1998, A\&A, 336, 231

\bibitem[\protect\citeauthoryear{Casares, Mouchet, Martinez-Pais \&
  Harlaftis}{Casares et~al.}{1996}]{casares96}
Casares J.,  Mouchet M.,  Martinez-Pais I.~G.,    Harlaftis E.~T.,  1996,
  MNRAS, 282, 182

\bibitem[\protect\citeauthoryear{Chanan, Middleditch \& Nelson}{Chanan
  et~al.}{1976}]{chanan76}
Chanan G.~A.,  Middleditch J.,    Nelson J.~E.,  1976, ApJ, 208, 512

\bibitem[\protect\citeauthoryear{Claret}{Claret}{1998}]{claret98}
Claret A.,  1998, A\&A, 335, 647

\bibitem[\protect\citeauthoryear{Collier~Cameron}{Collier~Cameron}{1992}]{came%
ron92}
Collier~Cameron A.,  1992, in Byrne P.~B.,  Mullan D.~J.,  eds, Surface
  Inhomogeneities on Late-type Stars Springer Verlag, p.~33

\bibitem[\protect\citeauthoryear{Collier~Cameron}{Collier~Cameron}{2001}]{came%
ron01}
Collier~Cameron A.,  2001, in Boffin H.,  Steeghs D.,  eds, Astrotomography:
  Indirect Imaging Methods in Observational Astronomy Springer-Verlag Lecture
  Notes in Physics, Berlin, p.~183

\bibitem[\protect\citeauthoryear{Collier~Cameron \& Unruh}{Collier~Cameron \&
  Unruh}{1994}]{cameron94}
Collier~Cameron A.,  Unruh Y.~C.,  1994, MNRAS, 269, 814

\bibitem[\protect\citeauthoryear{Crawford \& Kraft}{Crawford \&
  Kraft}{1956}]{crawford56}
Crawford J.~A.,  Kraft R.~P.,  1956, ApJ, 123, 44

\bibitem[\protect\citeauthoryear{Davey \& Smith}{Davey \&
  Smith}{1992}]{davey92}
Davey S.,  Smith R.~C.,  1992, MNRAS, 257, 476

\bibitem[\protect\citeauthoryear{Davey \& Smith}{Davey \&
  Smith}{1996}]{davey96}
Davey S.,  Smith R.~C.,  1996, MNRAS, 280, 481

\bibitem[\protect\citeauthoryear{de Martino, Matt, G{\"{a}}nsicke, Silvotti,
  Bonnet-Bidaud \& Mouchet}{de~Martino et~al.}{2002}]{demartino02}
de Martino D.,  Matt G.,  G{\"{a}}nsicke B.~T.,  Silvotti R.,  Bonnet-Bidaud
  J.~M.,    Mouchet M.,  2002, A\&A, 396, 213

\bibitem[\protect\citeauthoryear{Dhillon \& Watson}{Dhillon \&
  Watson}{2001}]{dhillon01}
Dhillon V.~S.,  Watson C.~A.,  2001, in Boffin H.,  Steeghs D.,  eds,
  Astrotomography: Indirect Imaging Methods in Observational Astronomy
  Springer-Verlag Lecture Notes in Physics, Berlin, p.~94

\bibitem[\protect\citeauthoryear{Donati}{Donati}{1999}]{donati99b}
Donati J.-F.,  1999, MNRAS, 302, 457

\bibitem[\protect\citeauthoryear{Donati, Semel, Carter, Rees \&
  Collier~Cameron}{Donati et~al.}{1997}]{donati97a}
Donati J.-F.,  Semel M.,  Carter B.~D.,  Rees D.~E.,    Collier~Cameron A.,
  1997, MNRAS, 291, 658

\bibitem[\protect\citeauthoryear{Echevarr\a'{\i}a, Smith, Pineda, Costero,
  Zharikov, Michel \& Spruit}{Echevarr\a'{\i}a et~al.}{2005}]{echevarria05}
Echevarr\a'{\i}a J.,  Smith R.~C.,  Pineda L.,  Costero R.,  Zharikov S.,
  Michel R.,    Spruit H.,  2005, MNRAS, to be submitted

\bibitem[\protect\citeauthoryear{Efron}{Efron}{1979}]{efron79}
Efron B.,  1979, Annals of Statistics, 7, 1

\bibitem[\protect\citeauthoryear{Efron \& Tibshirani}{Efron \&
  Tibshirani}{1993}]{efron93}
Efron B.,  Tibshirani R.~J.,  1993, An Introduction to the Bootstrap.
Chapman \& Hall, New York

\bibitem[\protect\citeauthoryear{Eracleous \& Horne}{Eracleous \&
  Horne}{1996}]{eracleous96}
Eracleous M.,  Horne K.~D.,  1996, ApJ, 471, 427

\bibitem[\protect\citeauthoryear{Granzer, Sch{\"{u}}ssler, Caligari \&
  Strassmeier}{Granzer et~al.}{2000}]{granzer00}
Granzer T.,  Sch{\"{u}}ssler M.,  Caligari P.,    Strassmeier K.~G.,  2000,
  A\&A, 355, 1087

\bibitem[\protect\citeauthoryear{Gray}{Gray}{1992}]{gray92}
Gray D.~F.,  1992, The Observations and Analysis of Stellar Photospheres.
Wiley-Interscience, New York

\bibitem[\protect\citeauthoryear{Hatzes \& Vogt}{Hatzes \&
  Vogt}{1992}]{hatzes92}
Hatzes A.~P.,  Vogt S.~S.,  1992, MNRAS, 258, 387

\bibitem[\protect\citeauthoryear{Hatzes, Vogt, Ramseyer \& Misch}{Hatzes
  et~al.}{1996}]{hatzes96}
Hatzes A.~P.,  Vogt S.~S.,  Ramseyer T.~F.,    Misch A.,  1996, AJ, 469, 808

\bibitem[\protect\citeauthoryear{Hellier, O'Donoghue, Buckley \&
  Norton}{Hellier et~al.}{1990}]{hellier90}
Hellier C.,  O'Donoghue D.,  Buckley D.,    Norton A.,  1990, MNRAS, 242, 32

\bibitem[\protect\citeauthoryear{Henden \& Honeycutt}{Henden \&
  Honeycutt}{1995}]{henden95}
Henden A.~A.,  Honeycutt R.~K.,  1995, PASP, 107, 324

\bibitem[\protect\citeauthoryear{Hessman, G{\"{a}}nsicke \& Mattei}{Hessman
  et~al.}{2000}]{hessman00}
Hessman F.~V.,  G{\"{a}}nsicke B.~T.,    Mattei J.~A.,  2000, A\&A, 361, 952

\bibitem[\protect\citeauthoryear{Holzwarth \& Sch{\"{u}}ssler}{Holzwarth \&
  Sch{\"{u}}ssler}{2003}]{holzwarth03}
Holzwarth V.,  Sch{\"{u}}ssler M.,  2003, A\&A, 405, 303

\bibitem[\protect\citeauthoryear{Horne}{Horne}{1986}]{horne86a}
Horne K.,  1986, pasp, 98, 609

\bibitem[\protect\citeauthoryear{Jeffers, Barnes \& Collier~Cameron}{Jeffers
  et~al.}{2002}]{jeffers02}
Jeffers S.~V.,  Barnes J.~R.,    Collier~Cameron A.,  2002, MNRAS, 331, 666

\bibitem[\protect\citeauthoryear{King \& Cannizzo}{King \&
  Cannizzo}{1998}]{king98}
King A.~R.,  Cannizzo J.~K.,  1998, ApJ, 499, 348

\bibitem[\protect\citeauthoryear{King, Villarreal, Soderblom, Gulliver \&
  Adelman}{King et~al.}{2003}]{king03}
King J.~R.,  Villarreal A.~R.,  Soderblom D.~R.,  Gulliver A.~F.,    Adelman
  S.~J.,  2003, AJ, 125, 1980

\bibitem[\protect\citeauthoryear{Kraft}{Kraft}{1967}]{kraft67}
Kraft R.~P.,  1967, ApJ, 150, 551

\bibitem[\protect\citeauthoryear{Kupka, Piskunov, Ryabchikova, Stempels \&
  W.}{Kupka et~al.}{1999}]{kupka99}
Kupka F.,  Piskunov N.~E.,  Ryabchikova T.~A.,  Stempels H.~C.,    W. W.~W.,
  1999, A\&AS, 138, 119

\bibitem[\protect\citeauthoryear{Kupka, Ryabchikova, Piskunov, Stempels \&
  W.}{Kupka et~al.}{2000}]{kupka00}
Kupka F.~G.,  Ryabchikova T.~A.,  Piskunov N.~E.,  Stempels H.~C.,    W. W.~W.,
   2000, Baltic Astronomy, 9, 590

\bibitem[\protect\citeauthoryear{Lister, Collier~Cameron \& Bartus}{Lister
  et~al.}{1999}]{lister99}
Lister T.,  Collier~Cameron A.,    Bartus J.,  1999, MNRAS, 307, 685

\bibitem[\protect\citeauthoryear{Livio \& Pringle}{Livio \&
  Pringle}{1994}]{livio94}
Livio M.,  Pringle J.~E.,  1994, ApJ, 427, 956

\bibitem[\protect\citeauthoryear{Marsden, Waite, Carter \& Donati}{Marsden
  et~al.}{2005}]{marsden05}
Marsden S.~C.,  Waite I.~A.,  Carter B.~D.,    Donati J.~F.,  2005, MNRAS, 359,
  711

\bibitem[\protect\citeauthoryear{Mauche, Lee \& Kallman}{Mauche
  et~al.}{1997}]{mauche97}
Mauche C.~W.,  Lee P.~Y.,    Kallman T.~R.,  1997, ApJ, 477, 832

\bibitem[\protect\citeauthoryear{Mestel}{Mestel}{1968}]{mestel68}
Mestel L.,  1968, MNRAS, 138, 359

\bibitem[\protect\citeauthoryear{O'Neal, Saar \& Neff}{O'Neal
  et~al.}{1996}]{oneal96}
O'Neal D.,  Saar S.~H.,    Neff J.~E.,  1996, ApJ, 463, 766

\bibitem[\protect\citeauthoryear{O'Neal, Saar \& Neff}{O'Neal
  et~al.}{1998}]{oneal98}
O'Neal D.,  Saar S.~H.,    Neff J.~E.,  1998, ApJ, 501, L73

\bibitem[\protect\citeauthoryear{Pearson, Horne \& Skidmore}{Pearson
  et~al.}{2003}]{pearson03}
Pearson K.~K.,  Horne K.,    Skidmore W.,  2003, MNRAS, 338, 1067

\bibitem[\protect\citeauthoryear{Petit, Donati, Wade, Landstreet, Bagnulo,
  L{\"{u}}ftinger, Sigut, Shorlin, Strasser, Auri\`{e}re \& Oliveira}{Petit
  et~al.}{2004}]{petit04}
Petit P.,  Donati J.-F.,  Wade G.~A.,  Landstreet J.~D.,  Bagnulo S.,
  L{\"{u}}ftinger T.,  Sigut T. A.~A.,  Shorlin S. L.~S.,  Strasser S.,
  Auri\`{e}re M.,    Oliveira J.~M.,  2004, MNRAS, 348, 1175

\bibitem[\protect\citeauthoryear{Pringle}{Pringle}{1985}]{pringle85}
Pringle J.~E.,  1985, in Pringle J.~E.,  Wade R.~A.,  eds, , Interacting Binary
  Stars.
Cambridge University Press, Cambridge, p.~1

\bibitem[\protect\citeauthoryear{Pulkkinen, Brooke, Pelt \& Tuominen}{Pulkkinen
  et~al.}{1999}]{pulkkinen99}
Pulkkinen P.~J.,  Brooke J.,  Pelt J.,    Tuominen I.,  1999, A\&A, 341, L43

\bibitem[\protect\citeauthoryear{Reinsch \& Beuermann}{Reinsch \&
  Beuermann}{1994}]{reinsch94}
Reinsch K.,  Beuermann K.,  1994, A\&A, 282, 493

\bibitem[\protect\citeauthoryear{Rice \& Strassmeier}{Rice \&
  Strassmeier}{1998}]{rice98}
Rice J.~B.,  Strassmeier K.~G.,  1998, A\&A, 336, 972

\bibitem[\protect\citeauthoryear{Richman, Applegate \& Patterson}{Richman
  et~al.}{1994}]{richman94}
Richman H.~R.,  Applegate J.~H.,    Patterson J.,  1994, PASP, 106, 1075

\bibitem[\protect\citeauthoryear{Rutten}{Rutten}{1987}]{rutten87}
Rutten R. G.~M.,  1987, A\&A, 177, 131

\bibitem[\protect\citeauthoryear{Rutten \& Dhillon}{Rutten \&
  Dhillon}{1994}]{rutten94}
Rutten R. G.~M.,  Dhillon V.~S.,  1994, A\&A, 288, 773

\bibitem[\protect\citeauthoryear{Scharlemann}{Scharlemann}{1982}]{scharlemann8%
2}
Scharlemann E.~T.,  1982, ApJ, 253, 298

\bibitem[\protect\citeauthoryear{Schenker, King, Kolb, Wynn \& Zhang}{Schenker
  et~al.}{2002}]{schenker02}
Schenker K.,  King A.~R.,  Kolb U.,  Wynn G.~A.,    Zhang Z.,  2002, MNRAS,
  337, 1105

\bibitem[\protect\citeauthoryear{Schreiber, G{\"{a}}nsicke \&
  Hessman}{Schreiber et~al.}{2000}]{schreiber00}
Schreiber M.~R.,  G{\"{a}}nsicke B.~T.,    Hessman F.~V.,  2000, A\&A, 358, 221

\bibitem[\protect\citeauthoryear{Schrijver \& Title}{Schrijver \&
  Title}{2001}]{schrijver01}
Schrijver C.~J.,  Title A.~M.,  2001, AJ, 551, 1099

\bibitem[\protect\citeauthoryear{Sch{\"{u}}ssler \& Solanki}{Sch{\"{u}}ssler \&
  Solanki}{1992}]{schussler92}
Sch{\"{u}}ssler M.,  Solanki S.~K.,  1992, A\&A, 355, 1087

\bibitem[\protect\citeauthoryear{Shahbaz}{Shahbaz}{1998}]{shahbaz98}
Shahbaz T.,  1998, MNRAS, 298, 153

\bibitem[\protect\citeauthoryear{Skidmore, O'Brien, Horne, Gomer, Oke \&
  Pearson}{Skidmore et~al.}{2003}]{skidmore03}
Skidmore W.,  O'Brien K.,  Horne K.,  Gomer R.,  Oke J.~B.,    Pearson K.~J.,
  2003, MNRAS, 338, 1057

\bibitem[\protect\citeauthoryear{Skilling \& Bryan}{Skilling \&
  Bryan}{1984}]{skilling84}
Skilling J.,  Bryan R.~K.,  1984, MNRAS, 211, 111

\bibitem[\protect\citeauthoryear{Spruit \& Ritter}{Spruit \&
  Ritter}{1983}]{spruit83}
Spruit H.~C.,  Ritter H.,  1983, A\&A, 124, 267

\bibitem[\protect\citeauthoryear{Strassmeier \& Bartus}{Strassmeier \&
  Bartus}{2000}]{strassmeier00}
Strassmeier K.~G.,  Bartus J.,  2000, A\&A, 354, 537

\bibitem[\protect\citeauthoryear{Strassmeier \& Rice}{Strassmeier \&
  Rice}{1998}]{strassmeier98}
Strassmeier K.~G.,  Rice J.~B.,  1998, A\&A, 330, 685

\bibitem[\protect\citeauthoryear{Taam \& Spruit}{Taam \& Spruit}{1989}]{taam89}
Taam R.~E.,  Spruit H.~C.,  1989, ApJ, 345, 972

\bibitem[\protect\citeauthoryear{Tanzi, Chincarini \& Tarenghi}{Tanzi
  et~al.}{1981}]{tanzi81}
Tanzi E.~G.,  Chincarini G.,    Tarenghi M.,  1981, PASP, 93, 68

\bibitem[\protect\citeauthoryear{Unruh \& Collier~Cameron}{Unruh \&
  Collier~Cameron}{1997}]{unruh97}
Unruh Y.~C.,  Collier~Cameron A.,  1997, MNRAS, 290, L37

\bibitem[\protect\citeauthoryear{Unruh, Collier~Cameron \& Cutispoto}{Unruh
  et~al.}{1995}]{unruh95b}
Unruh Y.~C.,  Collier~Cameron A.,    Cutispoto G.,  1995, MNRAS, 277, 1145

\bibitem[\protect\citeauthoryear{Vogt, Hatzes, Misch \& K{\"{u}}rster}{Vogt
  et~al.}{1999}]{vogt99}
Vogt S.~S.,  Hatzes A.~P.,  Misch A.~A.,    K{\"{u}}rster M.,  1999, ApJS, 121,
  547

\bibitem[\protect\citeauthoryear{Vogt \& Penrod}{Vogt \& Penrod}{1983}]{vogt83}
Vogt S.~S.,  Penrod G.~D.,  1983, PASP, 95, 565

\bibitem[\protect\citeauthoryear{Wade, Donati, Landstreet \& Shorlin}{Wade
  et~al.}{2000}]{wade00}
Wade G.~A.,  Donati J.-F.,  Landstreet J.~D.,    Shorlin S. L.~S.,  2000,
  MNRAS, 313, 823

\bibitem[\protect\citeauthoryear{Wade}{Wade}{1982}]{wade82}
Wade R.~A.,  1982, AJ, 87, 1558

\bibitem[\protect\citeauthoryear{Watson \& Dhillon}{Watson \&
  Dhillon}{2001}]{watson01a}
Watson C.~A.,  Dhillon V.~S.,  2001, MNRAS, 326, 67

\bibitem[\protect\citeauthoryear{Watson, Dhillon, Rutten \& Schwope}{Watson
  et~al.}{2003}]{watson03}
Watson C.~A.,  Dhillon V.~S.,  Rutten R. G.~M.,    Schwope A.~D.,  2003, MNRAS,
  341, 129

\bibitem[\protect\citeauthoryear{Webb, Naylor \& Jeffries}{Webb
  et~al.}{2002}]{webb02}
Webb N.~A.,  Naylor T.,    Jeffries R.~D.,  2002, ApJ, 568, 45

\bibitem[\protect\citeauthoryear{Welsh, Horne \& Gomer}{Welsh
  et~al.}{1995}]{welsh95}
Welsh W.~F.,  Horne K.,    Gomer R.,  1995, MNRAS, 275, 649

\bibitem[\protect\citeauthoryear{Welsh, Horne \& Oke}{Welsh
  et~al.}{1993}]{welsh93}
Welsh W.~F.,  Horne K.,    Oke J.~B.,  1993, ApJ, 406, 229

\bibitem[\protect\citeauthoryear{Wynn, King \& Horne}{Wynn
  et~al.}{1997}]{wynn97}
Wynn G.~A.,  King A.~R.,    Horne K.,  1997, MNRAS, 286, 436

\end{thebibliography}

\end{document}